\documentclass{aastex62}
\usepackage{amsmath}
\usepackage{amssymb}
\usepackage{graphicx}
\usepackage{longtable}
\usepackage{xcolor}
\usepackage{threeparttable}

\newcommand{\mytilde}{\raise.17ex\hbox{$\scriptstyle\mathtt{\sim}$}}

\newcommand{\TESS}{\textit{TESS }}
  \def \teff {$T_{\mathrm{eff}}$}

 \newcommand{\rplan}{3.45$^{+0.16}_{-0.12}$~$R_{\oplus}$ }
\newcommand{\mplan}{40.8$^{+2.4}_{-2.5}$~$M_{\oplus}$ }
\newcommand{\denplan}{$5.5\pm0.8$~gcm$^{-3}$ }
\newcommand{\pplan}{0.7655240~$\pm$~0.0000027~d }

 \newcommand{\rstar}{0.919$^{+0.031}_{-0.022}$~$R_{\odot}$ }
\newcommand{\mstar}{0.929~$\pm$~0.023~$M_{\odot}$ }
\newcommand{\agestar}{6.7$^{+2.8}_{-2.4}$~Gyr }

\begin{document}

\title{A remnant planetary core in the hot Neptunian desert}

\author[0000-0002-5080-4117]{David J. Armstrong}
\altaffiliation{STFC Ernest Rutherford Fellow}
\affil{Centre for Exoplanets and Habitability, University of Warwick, Gibbet Hill Road, Coventry, CV4 7AL, UK}
\affil{Department of Physics, University of Warwick, Gibbet Hill Road, Coventry, CV4 7AL, UK}

\author[0000-0001-6622-1250]{Th\'{e}o A. Lopez}
\affil{Aix Marseille Univ, CNRS, CNES, LAM, Marseille, France} 

\author[0000-0002-0601-6199]{Vardan Adibekyan}
\affil{Instituto de Astrof\'isica e Ci\^{e}ncias do Espa\c co, Universidade do Porto, CAUP, Rua das Estrelas, 4150-762 Porto, Portugal}

\author[0000-0002-0364-937X]{Richard A.\ Booth}
\affil{Institute of Astronomy, Madingley Road, Cambridge CB3 0HA, UK}

\author{Edward M. Bryant}
\affil{Centre for Exoplanets and Habitability, University of Warwick, Gibbet Hill Road, Coventry, CV4 7AL, UK}
\affil{Department of Physics, University of Warwick, Gibbet Hill Road, Coventry, CV4 7AL, UK}

\author[0000-0001-6588-9574]{Karen A.\ Collins}
\affiliation{Center for Astrophysics ${\rm \mid}$ Harvard {\rm \&} Smithsonian, 60 Garden Street, Cambridge, MA 02138, USA}

\author[0000-0002-8811-1914]{Alexandre Emsenhuber}
\affil{Lunar and Planetary Laboratory, University of Arizona, 1629 E. University Blvd., Tucson, AZ 85721, USA}
\affil{Physikalisches Institut, University of Bern, Gesellschaftsstrasse 6, 3012 Bern, Switzerland}

\author[0000-0003-0918-7484]{Chelsea X. Huang}
\affiliation{Department of Physics and Kavli Institute for Astrophysics and Space Research, Massachusetts Institute of Technology, Cambridge, MA 02139, USA}

\author{George W. King}
\affil{Centre for Exoplanets and Habitability, University of Warwick, Gibbet Hill Road, Coventry, CV4 7AL, UK}
\affil{Department of Physics, University of Warwick, Gibbet Hill Road, Coventry, CV4 7AL, UK}

\author{Jorge Lillo-box}
\affiliation{Depto. de Astrof\'isica, Centro de Astrobiolog\'ia (CSIC-INTA), ESAC campus, 28692 Villanueva de la Ca\~nada (Madrid), Spain}

\author[0000-0001-6513-1659]{Jack J. Lissauer}
\affiliation{NASA Ames Research Center, Moffett Field, CA, 94035, USA}

\author[0000-0003-0593-1560]{Elisabeth~C.~Matthews}
\affiliation{Department of Physics and Kavli Institute for Astrophysics and Space Research, Massachusetts Institute of Technology, Cambridge, MA 02139, USA}

\author{Olivier Mousis}
\affil{Aix Marseille Univ, CNRS, CNES, LAM, Marseille, France} 

\author[0000-0002-5254-2499]{Louise D. Nielsen}
\affil{Observatoire Astronomique de l'Universit\'e de Gen\'eve, 51 Chemin des Maillettes, 1290 Versoix, Switzerland}

\author[0000-0002-4047-4724]{Hugh Osborn}
\affil{Aix Marseille Univ, CNRS, CNES, LAM, Marseille, France}

\author{Jon Otegi}
\affil{Observatoire Astronomique de l'Universit\'e de Gen\'eve, 51 Chemin des Maillettes, 1290 Versoix, Switzerland}
\affil{Institute for Computational Science, University of Zurich,
              Winterthurerstr. 190, CH-8057 Zurich, Switzerland}

\author{Nuno C. Santos}
\affil{Instituto de Astrof\'isica e Ci\^{e}ncias do Espa\c co, Universidade do Porto, CAUP, Rua das Estrelas, 4150-762 Porto, Portugal}
\affil{Departamento de F\'isica e Astronomia, Faculdade de Ci\^encias, Universidade do Porto, Rua do Campo Alegre, 4169-007 Porto, Portugal}

\author[0000-0001-9047-2965]{S\'{e}rgio G. Sousa}
\affil{Instituto de Astrof\'isica e Ci\^{e}ncias do Espa\c co, Universidade do Porto, CAUP, Rua das Estrelas, 4150-762 Porto, Portugal}

\author[0000-0002-3481-9052]{Keivan G.\ Stassun}
\affiliation{Vanderbilt University, Department of Physics \& Astronomy, 6301 Stevenson Center Ln., Nashville, TN 37235, USA}
\affiliation{Fisk University, Department of Physics, 1000 18th Ave. N., Nashville, TN 37208, USA}

\author{Dimitri Veras}
\altaffiliation{STFC Ernest Rutherford Fellow}
\affil{Centre for Exoplanets and Habitability, University of Warwick, Gibbet Hill Road, Coventry, CV4 7AL, UK}
\affil{Department of Physics, University of Warwick, Gibbet Hill Road, Coventry, CV4 7AL, UK}

\author[0000-0002-0619-7639]{Carl Ziegler}
\affil{Dunlap Institute for Astronomy and Astrophysics, University of Toronto, 50 St. George Street, Toronto, Ontario M5S 3H4, Canada}


\author{Jack~S.~Acton}
\affil{School of Physics and Astronomy, University of Leicester, Leicester LE1 7RH, UK}

\author[0000-0003-3208-9815]{Jose M. Almenara}
\affiliation{Univ. Grenoble Alpes, CNRS, IPAG, 38000 Grenoble, France}

\author[0000-0001-7416-7522]{David~R.~Anderson}
\affil{Centre for Exoplanets and Habitability, University of Warwick, Gibbet Hill Road, Coventry, CV4 7AL, UK}
\affil{Department of Physics, University of Warwick, Gibbet Hill Road, Coventry, CV4 7AL, UK}

\author[0000-0002-5971-9242]{David Barrado}
\affiliation{Depto. de Astrof\'isica, Centro de Astrobiolog\'ia (CSIC-INTA), ESAC campus 28692 Villanueva de la Ca\~nada (Madrid), Spain}

\author{Susana C.C. Barros}
\affil{Instituto de Astrof\'isica e Ci\^{e}ncias do Espa\c co, Universidade do Porto, CAUP, Rua das Estrelas, 4150-762 Porto, Portugal}

\author[0000-0001-6023-1335]{Daniel Bayliss}
\affil{Centre for Exoplanets and Habitability, University of Warwick, Gibbet Hill Road, Coventry, CV4 7AL, UK}
\affil{Department of Physics, University of Warwick, Gibbet Hill Road, Coventry, CV4 7AL, UK}

\author{Claudia Belardi}
\affil{School of Physics and Astronomy, University of Leicester, Leicester LE1 7RH, UK}

\author{Francois Bouchy}
\affil{Observatoire Astronomique de l'Universit\'e de Gen\'eve, 51 Chemin des Maillettes, 1290 Versoix, Switzerland}

\author{C\'{e}sar Brice\~{n}o}
\affiliation{Cerro Tololo Inter-American Observatory, Casilla 603, La Serena, Chile} 

\author[0000-0002-7704-0153]{Matteo Brogi}
\affil{Centre for Exoplanets and Habitability, University of Warwick, Gibbet Hill Road, Coventry, CV4 7AL, UK}
\affil{Department of Physics, University of Warwick, Gibbet Hill Road, Coventry, CV4 7AL, UK}
\affil{INAF - Osservatorio Astrofisico di Torino, Via Osservatorio 20, 10025, Pino Torinese, Italy}

\author[0000-0003-1098-2442]{David J.~A.~Brown}
\affil{Centre for Exoplanets and Habitability, University of Warwick, Gibbet Hill Road, Coventry, CV4 7AL, UK}
\affil{Department of Physics, University of Warwick, Gibbet Hill Road, Coventry, CV4 7AL, UK}

\author{Matthew~R.~Burleigh}
\affil{School of Physics and Astronomy, University of Leicester, Leicester LE1 7RH, UK}

\author{Sarah~L.~Casewell}
\affil{School of Physics and Astronomy, University of Leicester, Leicester LE1 7RH, UK}

\author{Alexander~Chaushev}
\affil{Center for Astronomy and Astrophysics, TU Berlin, Hardenbergstr. 36, D-10623 Berlin, Germany}

\author[0000-0002-5741-3047]{David~ R.~Ciardi}
\affiliation{Caltech/IPAC-NASA Exoplanet Science Institute, 770 S. Wilson Avenue, Pasadena, CA 91106, USA}

\author[0000-0003-2781-3207]{Kevin I.\ Collins}
\affiliation{George Mason University, 4400 University Drive, Fairfax, VA, 22030 USA}

\author[0000-0001-8020-7121]{Knicole D. Col\'{o}n}
\affiliation{NASA Goddard Space Flight Center, Exoplanets and Stellar Astrophysics Laboratory (Code 667), Greenbelt, MD 20771, USA}
 
\author[0000-0002-8824-9956]{Benjamin F. Cooke}
\affil{Centre for Exoplanets and Habitability, University of Warwick, Gibbet Hill Road, Coventry, CV4 7AL, UK}
\affil{Department of Physics, University of Warwick, Gibbet Hill Road, Coventry, CV4 7AL, UK}

\author{Ian~J.~M.~Crossfield}
\affiliation{Department of Physics and Kavli Institute for Astrophysics and Space Research, Massachusetts Institute of Technology, Cambridge, MA 02139, USA}

\author[0000-0001-9289-5160]{Rodrigo~F.~D\'iaz}
\affiliation{Universidad de Buenos Aires, Facultad de Ciencias Exactas y Naturales. Buenos Aires, Argentina.}
\affiliation{CONICET - Universidad de Buenos Aires. Instituto de Astronom\'ia y F\'isica del Espacio (IAFE). Buenos Aires, Argentina.}

\author{Magali Deleuil}
\affil{Aix Marseille Univ, CNRS, CNES, LAM, Marseille, France} 

\author[0000-0003-4434-2195]{Elisa Delgado Mena}
\affil{Instituto de Astrof\'isica e Ci\^{e}ncias do Espa\c co, Universidade do Porto, CAUP, Rua das Estrelas, 4150-762 Porto, Portugal}

\author[0000-0001-7918-0355]{Olivier~D.~S. Demangeon}
\affil{Instituto de Astrof\'isica e Ci\^{e}ncias do Espa\c co, Universidade do Porto, CAUP, Rua das Estrelas, 4150-762 Porto, Portugal}

\author{Caroline Dorn}
\affil{Institute for Computational Science, University of Zurich,
              Winterthurerstr. 190, CH-8057 Zurich, Switzerland}
              
\author[0000-0002-9332-2011]{Xavier Dumusque}
\affil{Observatoire Astronomique de l'Universit\'e de Gen\'eve, 51 Chemin des Maillettes, 1290 Versoix, Switzerland}

\author[0000-0003-4096-0594]{Philipp Eigm\"uller}
\affil{Institute of Planetary Research, German Aerospace Center, Rutherfordstrasse 2, 12489 Berlin, Germany}

\author[0000-0002-9113-7162]{Michael Fausnaugh}
\affiliation{Department of Physics and Kavli Institute for Astrophysics and Space Research, Massachusetts Institute of Technology, Cambridge, MA 02139, USA}

\author{Pedro Figueira}
\affiliation{European Southern Observatory, Alonso de Cordova 3107, Vitacura, Santiago, Chile}
\affiliation{Instituto de Astrof\'isica e Ci\^{e}ncias do Espa\c co, Universidade do Porto, CAUP, Rua das Estrelas, 4150-762 Porto, Portugal}

\author{Tianjun Gan}
\affiliation{Department of Astronomy and Tsinghua Centre for Astrophysics, Tsinghua University, Beijing 100084, China}

\author[0000-0001-9552-3709]{Siddharth Gandhi}
\affil{Department of Physics, University of Warwick, Gibbet Hill Road, Coventry, CV4 7AL, UK}

\author[0000-0002-4259-0155]{Samuel Gill}
\affil{Centre for Exoplanets and Habitability, University of Warwick, Gibbet Hill Road, Coventry, CV4 7AL, UK}
\affil{Department of Physics, University of Warwick, Gibbet Hill Road, Coventry, CV4 7AL, UK}

\author{Michael~R.~Goad}
\affil{School of Physics and Astronomy, University of Leicester, Leicester LE1 7RH, UK}

\author[0000-0002-3164-9086]{Maximilian~N.~G{\"u}nther} 
\affiliation{Department of Physics and Kavli Institute for Astrophysics and Space Research, Massachusetts Institute of Technology, Cambridge, MA 02139, USA}
\altaffiliation{Juan Carlos Torres Fellow}

\author[0000-0001-5555-2652]{Ravit Helled}
\affil{Institute for Computational Science, University of Zurich,
              Winterthurerstr. 190, CH-8057 Zurich, Switzerland}
              
\author[0000-0002-0417-1902]{Saeed Hojjatpanah} 
\affiliation{Instituto de Astrof\'isica e Ci\^{e}ncias do Espa\c co, Universidade do Porto, CAUP, Rua das Estrelas, 4150-762 Porto, Portugal}
\affiliation{Departamento de F\'isica e Astronomia, Faculdade de Ci\^encias, Universidade do Porto, Rua do Campo Alegre, 4169-007 Porto, Portugal}

\author[0000-0002-2532-2853]{Steve~B.~Howell}
\affil{NASA Ames Research Center, Moffett Field, CA 94035, USA}

\author{James Jackman}
\affil{Centre for Exoplanets and Habitability, University of Warwick, Gibbet Hill Road, Coventry, CV4 7AL, UK}
\affil{Department of Physics, University of Warwick, Gibbet Hill Road, Coventry, CV4 7AL, UK}

\author{James S. Jenkins}
\affiliation{Departamento de Astronom\'ia, Universidad de Chile, Camino el Observatorio 1515, Casilla 36-D, Las Condes, Santiago, Chile}
\affiliation{Centro de Astrof\'isica y Tecnolog\'ias Afines (CATA), Casilla 36-D, Santiago, Chile}

\author[0000-0002-4715-9460]{Jon M. Jenkins}
\affiliation{NASA Ames Research Center, Moffett Field, CA, 94035, USA}

\author[0000-0002-4625-7333]{Eric L. N. Jensen}
\affiliation{Dept.\ of Physics \& Astronomy, Swarthmore College, Swarthmore PA 19081, USA}

\author[0000-0001-6831-7547]{Grant M. Kennedy}
\altaffiliation{Royal Society University Research Fellow}
\affil{Centre for Exoplanets and Habitability, University of Warwick, Gibbet Hill Road, Coventry, CV4 7AL, UK}
\affil{Department of Physics, University of Warwick, Gibbet Hill Road, Coventry, CV4 7AL, UK}

\author[0000-0001-9911-7388]{David~W.~Latham}
\affiliation{Harvard-Smithsonian Center for Astrophysics, 60 Garden St, Cambridge, MA 02138, USA}

\author{Nicholas Law}
\affiliation{Department of Physics and Astronomy, University of North Carolina at Chapel Hill, Chapel Hill, NC 27599-3255, USA}

\author[0000-0001-9699-1459]{Monika~Lendl}
\affil{Observatoire Astronomique de l'Universit\'e de Gen\'eve, 51 Chemin des Maillettes, 1290 Versoix, Switzerland}
\affiliation{Space Research Institute, Austraian Academy of Sciences, Schmiedlstr. 6, 8042 Graz, Austria}

\author{Michael Lozovsky}
\affil{Institute for Computational Science, University of Zurich,
              Winterthurerstr. 190, CH-8057 Zurich, Switzerland}
              
\author[0000-0003-3654-1602]{Andrew~W.~Mann}
\affiliation{Department of Physics and Astronomy, University of North Carolina at Chapel Hill, Chapel Hill, NC 27599-3255, USA}

\author{Maximiliano~Moyano}
\affiliation{Instituto de Astronom\'ia, Universidad Cat\'olica del Norte, Angamos 0610, 1270709, Antofagasta, Chile.}

\author[0000-0003-1631-4170]{James McCormac}
\affil{Centre for Exoplanets and Habitability, University of Warwick, Gibbet Hill Road, Coventry, CV4 7AL, UK}
\affil{Department of Physics, University of Warwick, Gibbet Hill Road, Coventry, CV4 7AL, UK}

\author{Farzana Meru}
\altaffiliation{Royal Society Dorothy Hodgkin Fellow}
\affil{Centre for Exoplanets and Habitability, University of Warwick, Gibbet Hill Road, Coventry, CV4 7AL, UK}
\affil{Department of Physics, University of Warwick, Gibbet Hill Road, Coventry, CV4 7AL, UK}

\author[0000-0002-1013-2811]{Christoph Mordasini}
\affil{Physikalisches Institut, University of Bern, Gesellschaftsstrasse 6, 3012 Bern, Switzerland}

\author{Ares Osborn}
\affil{Centre for Exoplanets and Habitability, University of Warwick, Gibbet Hill Road, Coventry, CV4 7AL, UK}
\affil{Department of Physics, University of Warwick, Gibbet Hill Road, Coventry, CV4 7AL, UK}

\author{Don Pollacco}
\affil{Centre for Exoplanets and Habitability, University of Warwick, Gibbet Hill Road, Coventry, CV4 7AL, UK}
\affil{Department of Physics, University of Warwick, Gibbet Hill Road, Coventry, CV4 7AL, UK}

\author[0000-0002-3012-0316]{Didier Queloz}
\affiliation{Cavendish Laboratory
J J Thomson Avenue
Cambridge, CB3 0HE, UK}

\author{Liam Raynard}
\affil{School of Physics and Astronomy, University of Leicester, Leicester LE1 7RH, UK}

\author[0000-0003-2058-6662]{George~R.~Ricker}
\affiliation{Department of Physics and Kavli Institute for Astrophysics and Space Research, Massachusetts Institute of Technology, Cambridge, MA 02139, USA}

\author[0000-0002-4829-7101]{Pamela~Rowden}
\affiliation{School of Physical Sciences, The Open University, Milton Keynes MK7 6AA, UK}

\author[0000-0002-3586-1316]{Alexandre Santerne}
\affil{Aix Marseille Univ, CNRS, CNES, LAM, Marseille, France} 

\author{Joshua E. Schlieder}
\affiliation{Exoplanets and Stellar Astrophysics Laboratory, Code 667, NASA Goddard Space Flight Center, Greenbelt, MD 20771, USA}

\author[0000-0002-6892-6948]{S.~Seager}
\affiliation{Department of Physics and Kavli Institute for Astrophysics and Space Research, Massachusetts Institute of Technology, Cambridge, MA 02139, USA}
\affiliation{Department of Earth, Atmospheric and Planetary Sciences, Massachusetts Institute of Technology, Cambridge, MA 02139, USA}
\affiliation{Department of Aeronautics and Astronautics, MIT, 77 Massachusetts Avenue, Cambridge, MA 02139, USA}

\author[0000-0001-5401-8079]{Lizhou Sha}
\affiliation{Department of Physics and Kavli Institute for Astrophysics and Space Research, Massachusetts Institute of Technology, Cambridge, MA 02139, USA}

\author[0000-0001-5603-6895]{Thiam-Guan Tan}
\affiliation{Perth Exoplanet Survey Telescope, Perth, Western Australia}

\author{Rosanna~H.~Tilbrook}
\affil{School of Physics and Astronomy, University of Leicester, Leicester LE1 7RH, UK}

\author[0000-0002-8219-9505]{Eric Ting}
\affiliation{NASA Ames Research Center, Moffett Field, CA, 94035, USA}

\author{St\'{e}phane Udry}
\affil{Observatoire Astronomique de l'Universit\'e de Gen\'eve, 51 Chemin des Maillettes, 1290 Versoix, Switzerland}

\author[0000-0001-6763-6562]{Roland~Vanderspek}
\affiliation{Department of Physics and Kavli Institute for Astrophysics and Space Research, Massachusetts Institute of Technology, Cambridge, MA 02139, USA}

\author{Christopher A. Watson}
\affil{Astrophysics Research Centre, Queen's University Belfast, Belfast, BT7 1NN, UK}

\author[0000-0001-6604-5533]{Richard G. West}
\affil{Centre for Exoplanets and Habitability, University of Warwick, Gibbet Hill Road, Coventry, CV4 7AL, UK}
\affil{Department of Physics, University of Warwick, Gibbet Hill Road, Coventry, CV4 7AL, UK}

\author[0000-0002-7823-1090]{Paul A. Wilson}
\affil{Centre for Exoplanets and Habitability, University of Warwick, Gibbet Hill Road, Coventry, CV4 7AL, UK}
\affil{Department of Physics, University of Warwick, Gibbet Hill Road, Coventry, CV4 7AL, UK}

\author[0000-0002-4265-047X]{Joshua~N.~Winn}
\affiliation{Department of Astrophysical Sciences, Princeton University, 4 Ivy Lane, Princeton, NJ 08544, USA}

\author[0000-0003-1452-2240]{Peter Wheatley}
\affil{Centre for Exoplanets and Habitability, University of Warwick, Gibbet Hill Road, Coventry, CV4 7AL, UK}
\affil{Department of Physics, University of Warwick, Gibbet Hill Road, Coventry, CV4 7AL, UK}

\author{Jesus Noel Villasenor}
\affil{Department of Physics and Kavli Institute for Astrophysics and Space Research, Massachusetts Institute of Technology, Cambridge, MA 02139, USA}

\author{Jose I. Vines}
\affil{Departamento de Astronom\'ia, Universidad de Chile, Camino el Observatorio 1515, Casilla 36-D, Las Condes, Santiago, Chile}

\author[0000-0002-4142-1800]{Zhuchang Zhan}
\affil{Department of Earth, Atmospheric and Planetary Sciences, Massachusetts Institute of Technology, Cambridge, MA 02139, USA}

\begin{abstract}

The interiors of giant planets remain poorly understood. Even for the planets in the Solar System, difficulties in observation lead to major uncertainties in the properties of planetary cores.  Exoplanets that have undergone rare evolutionary pathways provide a new route to understanding planetary interiors. We present the discovery of TOI-849b, the remnant core of a giant planet, with a radius smaller than Neptune but an anomalously high mass $M_p=$\mplan and density of \denplan, similar to the Earth. Interior structure models suggest that any gaseous envelope of pure hydrogen and helium consists of no more than $3.9^{+0.8} _{-0.9}$\% of the total mass of the planet. TOI-849b transits a late G type star (T$_{\rm mag}=11.5$) with an orbital period of 18.4 hours, leading to an equilibrium temperature of 1800K. 
The planet's mass is larger than the theoretical threshold mass for runaway gas accretion. As such, the planet could have been a gas giant before undergoing extreme mass loss via thermal self-disruption or giant planet collisions, or it avoided substantial gas accretion, perhaps through gap opening or late formation. Photoevaporation rates cannot provide the mass loss required to reduce a Jupiter-like gas giant, but can remove a few $M_\oplus$ hydrogen and helium envelope on timescales of several Gyr, implying that any remaining atmosphere is likely to be enriched by water or other volatiles from the planetary interior. TOI-849b represents a unique case where material from the primordial core is left over from formation and available to study. 

\end{abstract}


\section{Main Text}

\begin{figure}
\resizebox{\hsize}{!}{\includegraphics{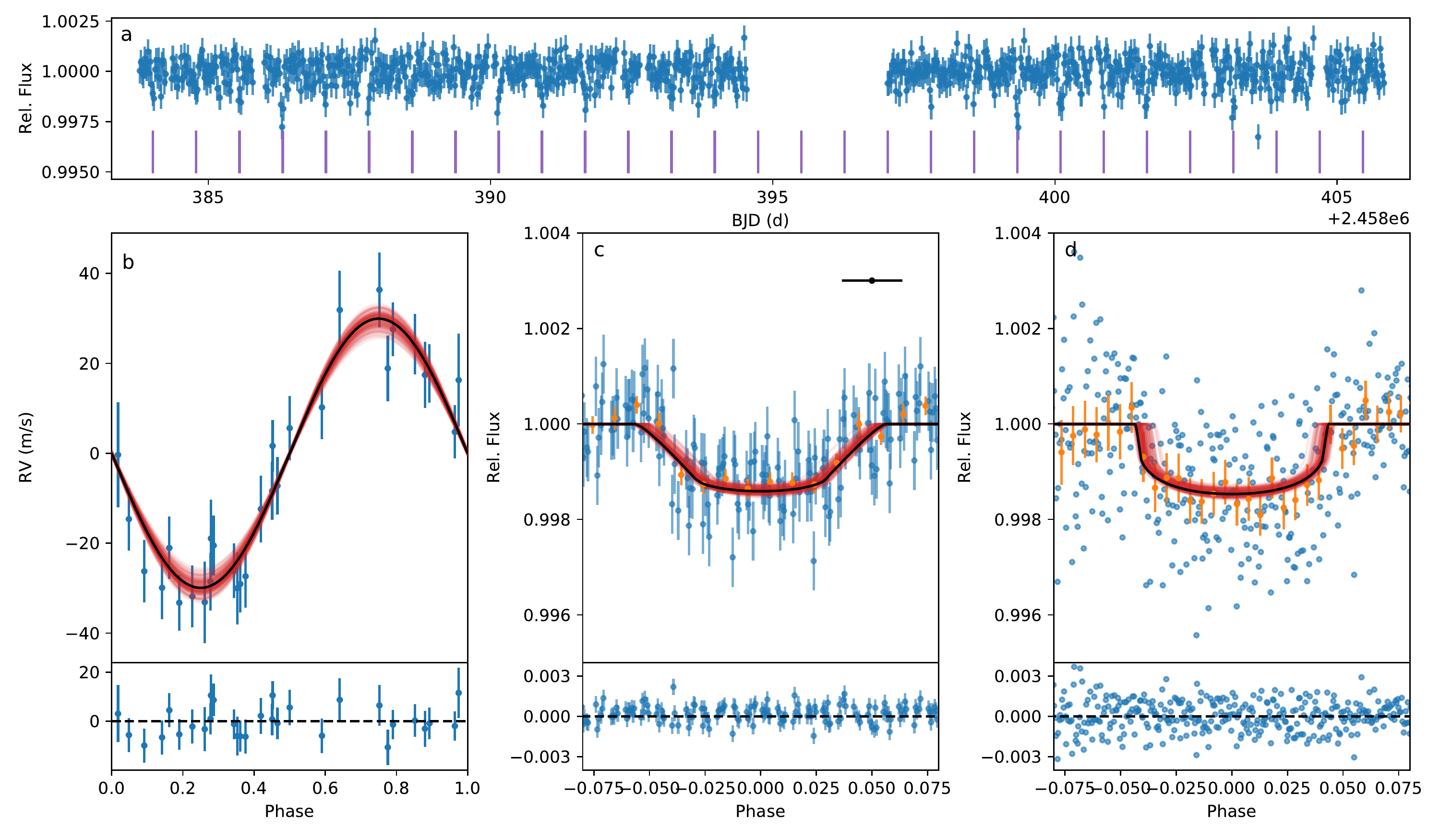}}
\caption{Best fitting model to the \TESS, HARPS and NGTS data. \textbf{a} \TESS lightcurve with transit times marked as vertical lines. \textbf{b} Phase-folded HARPS data and best fitting model in black, with residuals below. Several models randomly drawn from the MCMC chain are shown in red. \textbf{c} Phase-folded TESS 30-minute cadence data in blue, binned to 0.01 in phase in orange, with models as in b and residuals below. Horizontal error bar shows the TESS cadence. \textbf{d} Phase-folded NGTS data binned to 1 minute (blue) and to 0.01 in phase (orange). We plot the binned NGTS data to aid visualisation but fit to the full dataset. Model draws are shown as in b, with residuals below. The cadence is negligible at this scale. LCOGT data was also used and is shown in Supplementary Figure 1.}
\label{figbestfit}
\end{figure}

The \TESS mission \citep{2015JATIS...1a4003R} observed the V$_\textrm{mag}=12$ star TOI-849/TIC33595516 for 27 days during September and October 2018, leading to the detection of a candidate transiting planet. TOI-849 was observed at 30-minute cadence in the Full Frame Images, and was discovered using the MIT quick-look pipeline (see Methods). No signs of additional planets or stellar activity were seen in the photometry. Follow-up observations with the High Accuracy Radial velocity Planet Searcher (HARPS) spectrograph detected a large radial velocity signal, confirming the planet TOI-849b. Four additional transits were observed using the ground-based telescopes of the Next Generation Transit Survey \cite[NGTS,][]{{NGTS-2018}} and Las Cumbres Observatory Global Telescope \citep[LCOGT,][]{Brown:2013}, significantly improving the radius determination and ephemeris of the planet. A search of the Gaia Data Release 2 reveals no other sources closer than 39\arcsec, with the closest source 7.8 magnitudes fainter than TOI-849 in the G band \citep{2018A&A...616A...1G}. Additional high resolution imaging from SOAR, NaCo and AstraLux revealed no unresolved companion stars. We perform a joint fit to the data using the \texttt{PASTIS} software \citep{Diaz:2014kd,Santerne:2015bb} to extract planetary and stellar parameters, using the combined HARPS spectra to derive priors on the stellar parameters and calculate chemical abundances for the host star (see Methods). The best fit and data are shown in Figure \ref{figbestfit}.

\begin{figure}
\resizebox{\hsize}{!}{\includegraphics{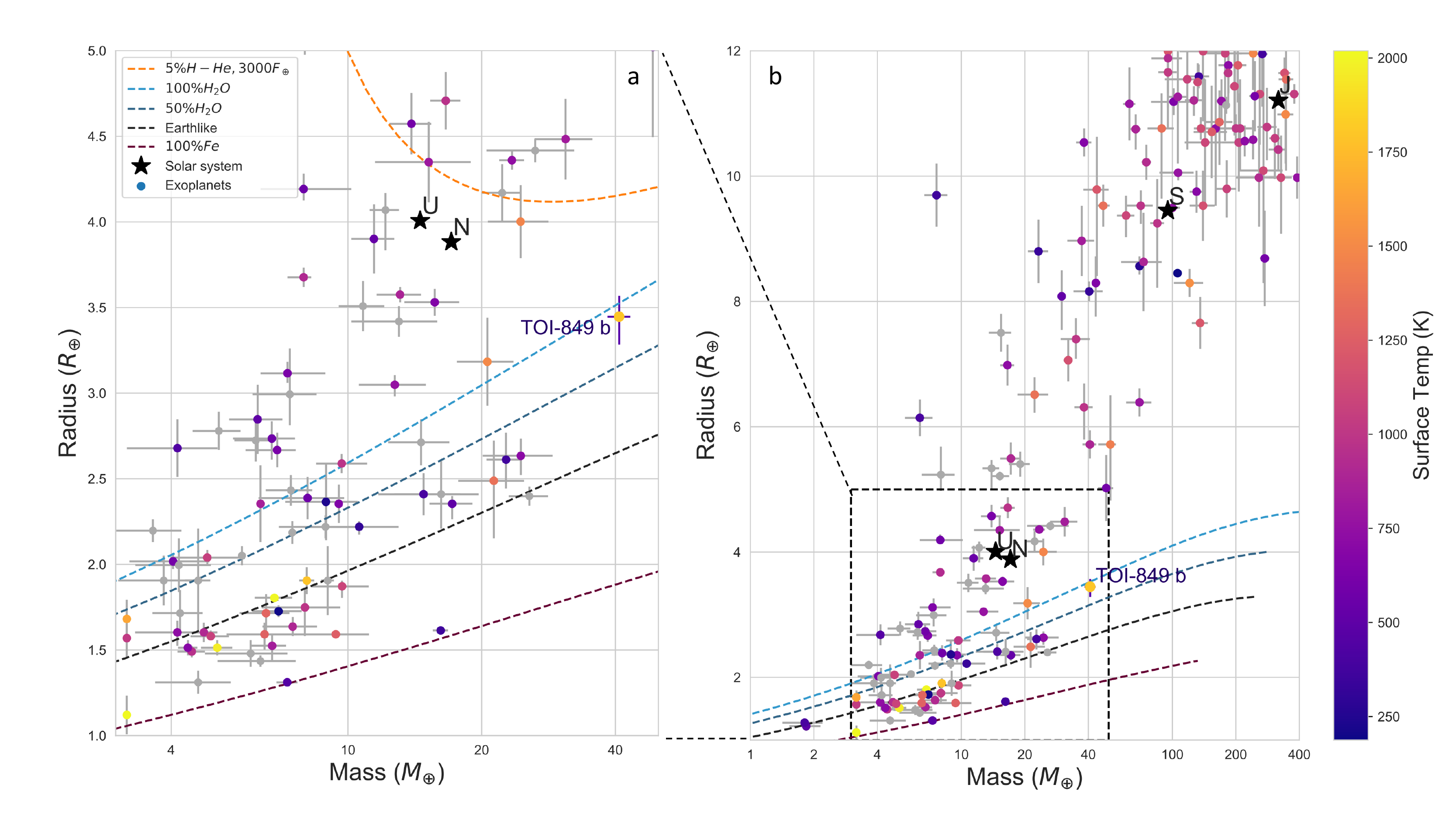}}
\caption{Mass-radius diagram of known exoplanets from the NASA exoplanet archive (https://exoplanetarchive.ipac.caltech.edu/) as of 17th October 2019. Planets are coloured by equilibrium temperature, where the information to calculate it is available on the archive, and are grey otherwise. Planets with mass determinations better than 4$\sigma$ are shown. Some planets where the source paper does not claim a mass determination, notably those from \citet{2014ApJS..210...25X}, were removed. Composition tracks from \citep{zeng2013detailed} are shown as dashed lines and defined in the figure legend, with an additional 5\%H-He track at an irradiation level similar to TOI-849b. \textbf{a} Zoom of panel \textbf{b}.}
\label{figMRplot}
\end{figure}

\begin{figure}
\resizebox{\hsize}{!}{\includegraphics{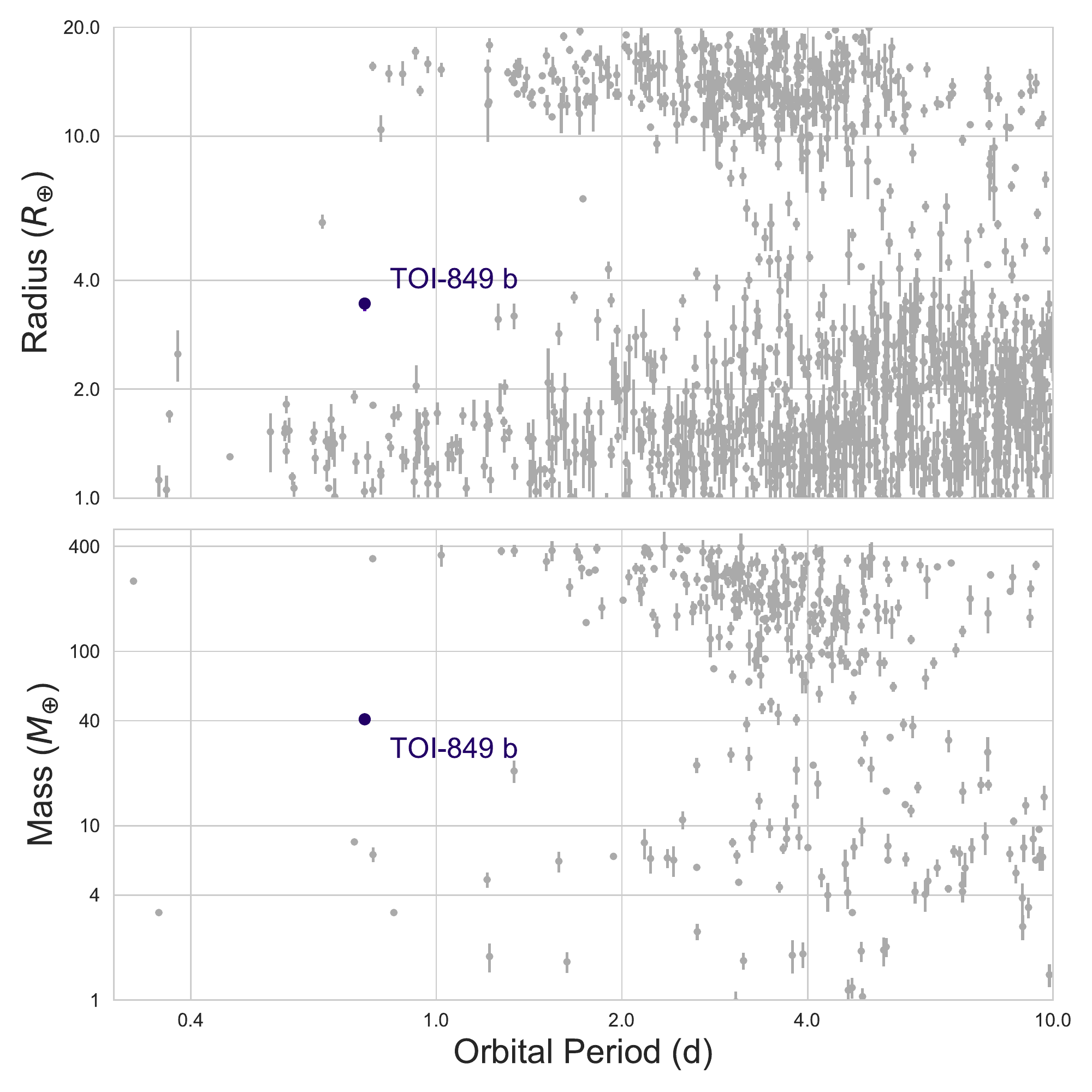}}
\caption{TOI-849b in the context of the Neptunian desert. Known exoplanets are plotted in grey and sourced from the NASA exoplanet archive (https://exoplanetarchive.ipac.caltech.edu/) as of 17th October 2019. Only planets with mass or radius determinations better than 4$\sigma$ are plotted. In both mass and radius TOI-849b lies in a sparsely populated region of the desert.}
\label{figNepDesert}
\end{figure}

TOI-849b has a mass of \mplan, nearly half the mass of Saturn. The planet's radius is \rplan and its mean density is \denplan, making it the densest Neptune-sized planet discovered to date (Figure \ref{figMRplot}). It has a sub-1d orbital period of \pplan, making it an 'ultra-short-period' (USP) planet and only the second such Neptune-sized object. The upper limit on its eccentricity is 0.08 at 95\% confidence. The radius, mass and period place TOI-849b in the middle of the hot Neptunian desert, a region of parameter space typically devoid of planets due to photoevaporation and tidal disruption \citep{2011ApJ...727L..44S,2013ApJ...763...12B,2016A&A...589A..75M,2018MNRAS.479.5012O} (Figure \ref{figNepDesert}). The host star TOI-849 is a late G dwarf with mass of \mstar, radius \rstar, and age \agestar. The close proximity of planet and star lead to an equilibrium temperature for the planet of 1800K, assuming an albedo of 0.3. The full set of derived parameters for the planet and star are given in Table \ref{MCMCprior}, and general stellar parameters in Table \ref{tabstellarproperties}.

TOI-849b represents a new frontier for interior structure models. The most widely used models of terrestrial planets are not valid for planets as massive as TOI-849b, because the properties of matter at such high central pressures remain highly uncertain. Furthermore, some compositional mixing is expected at these high pressures and temperatures \citep{Bodenheimer-18}, in contradiction of the usual assumption of distinct layers \citep[e.g.][]{Dorn-17A}. We build an internal structure model based on a modified version of \citet{Dorn-17A} (see Methods), accounting for some of these issues, but restrict our analysis to considering the limiting cases of a maximum and minimum possible hydrogen/helium (H/He) envelope under the layered structure assumption.  We calculate the maximum envelope mass by minimising the contribution of core, mantle and water, assuming the planet has the same [Fe/Si] ratio as has been observed for the photosphere of the host star. Under this model, the maximum envelope mass fraction is $3.9^{+0.8} _{-0.9}$\%. 

TOI-849b presents an interesting comparison to other relatively large planets with high metallicity. Two other recent planets discovered in the Neptunian desert are both expected to have a small envelope mass fraction. NGTS-4b \citep{West-19} has a period of 1.34d, mass of $20.6\pm3M_\oplus$ and radius of $3.18\pm0.26R_\oplus$, placing it on the pure water composition track on the M-R diagram, similar to TOI-849b but at much lower mass (Figure \ref{figMRplot}). LTT9779b (Jenkins et al, submitted), the only other USP Neptune known, has a period of 0.79d, mass of $29.3\pm0.8M_\oplus$ and radius of $4.59\pm0.23R_\oplus$. TOI-849b is more massive and of higher density than both these objects, implying it could be an extreme case of whatever formation process is populating the desert.

TOI-849b's large core mass and low envelope mass fraction challenge the traditional view of planet formation via core accretion, where planets with masses above a critical mass of \mytilde10--20$M_\oplus$ are expected to undergo runaway gas accretion within the protoplanetary disc \citep{Mizuno-78,Rafikov-06,Movshovitz-10,Lee-14,Piso-15}.
Why, then, does TOI-849b lack a massive gaseous envelope?
Apparently, the core somehow avoided runaway accretion, or else the planet was once
a gas giant which somehow lost most of its envelope. If runaway accretion proceeded to produce a giant planet, significant reduction in the original mass would be required to reach the present day state. HD149026b \citep{Sato-05} is a giant planet with mass $121\pm19M_\oplus$\citep{Stassun-17} thought to have a solid core with a mass of \mytilde50$M_\oplus$\citep{Fortney-06,Ikoma-06}, similar to TOI-849b. Starting from a planet like HD149026b, mass-loss of 60--70\% would be required to produce the present day TOI-849b. Considering the proximity of TOI-849b to its host star, one would expect some mass-loss to photoevaporation. The lifetime predicted mass-loss rate for a Jupiter-like planet is only a few percent, well below the required range (see Methods). For a planet like HD149026b the situation is less clear, and the lifetime mass removed depends critically on the assumptions made. We proceed to explore several formation pathways for TOI-849b.

Tidal disruption could cause mass loss of one--two orders of magnitude. The close proximity of a number of hot Jupiters to their tidal disruption radii \cite[e.g.][]{Delrez-16} and the fact that hot Jupiters are preferentially found around younger stars \citep{CollCam-18,Hamer-19} suggest that tidal disruption of hot Jupiters might be common. Although it appears they do not typically leave behind a remnant core, or such cores are short-lived \citep{2017AJ....154...60W}, as a rare higher mass object TOI-849b may be an unusual case. At the location of TOI-849b, tidal disruption would be expected for a Jupiter-mass planet with radius $>1.5$ Jupiter radii. An alternative, related pathway to substantial envelope loss is disruption via tidal thermalisation events, which can lead to mass loss of order one to two magnitudes. If TOI-849b reached its close orbit via high-eccentricity scattering by another planet in the system, energy build up in the planet's internal f-modes during tidal circularisation can approach significant fractions of the planet's internal binding energy and potentially lead to thermalisation events \citep{2019MNRAS.484.5645V,2019MNRAS.489.2941V}, which may remove envelope layers (see Methods). However, in either case it is unclear whether a giant planet could harbour a large enough core to leave behind a 40$M_\oplus$ remnant, because the gaseous envelope on top of a few $M_\oplus$ core causes planetesimals to be eroded in the envelope. The remaining solids must subsequently rain out to produce such a large core \citep{Iaroslavitz-07,Brouwers-18,Bodenheimer-18}.

Giant planet collisions provide another, intermediate way to produce planets similar to TOI-849b. The Bern planetary population synthesis models \citep{2018haex.bookE.143M} predict the existence of a small population of planets with similar masses and semi-major axes to TOI-849b (see Methods). In those models such planets were produced via giant planet collisions at the end of the migration phase, resulting in the ejection of the planetary envelope, and leaving no time for the remnant core to accrete further gas. In these scenarios, the cores reached an envelope mass fraction of a few tens of percent, before being reduced to Neptune size and ejecting the envelope through an impact. Such a scenario leaves a dense planetary core close to the host star. 

The alternative hypothesis is for TOI-849b to avoid runaway accretion, possibly through opening a gap in the protoplanetary disc, largely devoid of gas, before the planet accretes much envelope mass. Because the threshold mass required for a planet to open up a gap in a protoplanetary disc is sensitive to the disc scale-height, which is small close to the star, planets on close in orbits can more easily open a deep gap. A 40$M_\oplus$ planet like TOI-849b on a 0.1AU orbit would reduce the disc surface density at its location by a factor \mytilde10 \citep{Crida-06,Duffell-13,Kanagawa-15}. Recently, it has been argued that a reduction in gas accretion due to gap opening is required to resolve the fact that runaway gas accretion models tend to produce too many Jupiter mass planets and not enough sub-Saturn mass planets \citep{Lee-19}. Indeed, by reducing the accretion rate onto gap-opening planets \citet{Lee-19} are able to produce 40$M_\oplus$ planets at 0.1 AU with gas mass fractions below 10\% if the planets form late enough. In contrast to the tidal disruption pathway, reduced gas accretion should leave TOI-849b aligned with the stellar spin axis. Detecting or ruling out such alignment using measurements of the Rossiter-McLaughlin effect \citep{2018haex.bookE...2T}, as well as taking measurements of the atmospheric composition, may aid in distinguishing between the various formation scenarios.

In all cases, remaining hydrogen and helium envelope masses of a few percent could be removed over several Gyr by photoevaporation, given the planet's close orbit. We estimate the current mass-loss rate to be $9.5\times10^{-10}$M$_\oplus$\,yr$^{-1}$ (see Methods), implying an envelope mass of \mytilde 4\% could be removed in a few Gyr. As such, the question changes: where does TOI-849b's minor envelope come from? Given the high equilibrium temperature, we would expect to evaporate some ices to provide a secondary enriched atmosphere containing water and other volatiles. In these circumstances TOI-849b provides a unique target where the composition of a primordial planetary core could be studied by observing its atmospheric constituents, with for example the Hubble or upcoming James Webb Space Telescopes. 

TOI-849b's proximity to its host star, encouraging gap opening and increasing the role of photoevaporation, could explain why similar objects have not yet been found. Ultimately, however TOI-849b formed, the planet's large mass and low gas mass fraction will provide a stringent test of planet formation theory. TOI-849b gives us a glimpse at a core similar to those that exist at the centres of giant planets, exposed through an unlikely combination of inhibited accretion or mass-loss. Future observations may be able to directly observe the composition of that core by detecting evaporated material in the planetary atmosphere. TOI-849b is only the second published planet to populate the Neptunian desert, and is unique in its anomalously high density, pointing to a rare formation and evolution pathway.

\section{Methods}

\subsection{\TESS}
TOI-849 was observed in \TESS sector 3 (Sep 20 2018-Oct 18 2018), Camera 2 and CCD 3, with 30 min cadence on the Full Frame Images (FFIs). The calibrated FFIs available at MAST were produced by the \TESS Science Processing Operations Center (SPOC) \citep{Jenkins-16}. The candidate is detected by the MIT Quick Look pipeline \citep{2019AAS...23320908H} with a signal to noise of 18. The candidate exhibited consistent transit depth in the multi-aperture analysis and appeared to be on target in the difference image analysis. It passed all the vetting criteria set by the \TESS Science Office and was released as a \TESS Object of Interest.  

The aperture showing minimal scatter was found to be circular with a radius of 2.5 pixels, with the background determined on an annulus with a width of 3 pixels and an inner radius of 4 pixels. We reject outliers due to momentum dump using the quaternion time series provided by the spacecraft data. Further long time scale trends are removed using a B-spline based algorithm \citep{2014PASP..126..948V}. No significant evidence of photometric activity was observed. The lightcurve was further detrended to remove residual long term trends using a modified Savitzky-Golay filter as detailed in \citet{2015A&A...579A..19A}, whereby a sliding window is used to fit a 3-dimensional polynomial function to the data while ignoring outliers. Both flattening operations were carried out ignoring in-transit datapoints. Data before 2458383.78 BJD and after 2458405.77 BJD are masked because, during this time, the TESS operations team carried out several experiments on the attitude control system, causing the jitter profile to differ from normal. Data points between 2458394.54 BJD to 2458397.0 BJD are masked because of scattered light. The resulting lightcurve is shown in Figure \ref{figbestfit}.

\subsection{NGTS}
Two full transits of TOI-849 were observed on the nights UT 2019 August 08 and 2019 August 11 using the Next Generation Transit Survey \cite[NGTS; ][]{NGTS-2018} at ESOs Paranal Observatory in Chile, and are plotted in Figure \ref{figbestfit}. The NGTS facility consists of 12 fully robotic 20cm telescopes coupled to Andor iKon-L 936 cameras, each with an instantaneous field-of-view of 8 square degrees and a pixel scale of ~5\arcsec\ per pixel.  On both nights, 10 NGTS telescopes were used to simultaneously observe the transit. The photometric noise was found to be uncorrelated between the individual NGTS telescopes, and so we can combine the light curves to achieve ultra-high precision photometry for TOI-849. A total of 29654 images were obtained with an exposure time of 10 seconds, using the custom NGTS filter (520 - 890 nm). The observations were all obtained at an airmass z$<$2 and with photometric observing conditions. The telescope guiding was performed using the DONUTS auto-guiding algorithm \citep{DONUTS}, which provides sub-pixel level stability of the target position on the CCD. We do not require the use of flat fields during the image reduction, as a result of the high precision of the auto-guiding. This reduction was performed using a custom aperture photometry pipeline, in which the 100 best comparison stars were selected and ranked based on their proximity to the target star in the parameters of on-sky-separation, apparent magnitude, and colour. This large number of optimised comparison stars can be chosen because of the wide field-of-view of the NGTS telescopes, and again improves the precision of the NGTS light curves by reducing the presence of correlated noise.

\subsection{HARPS}
We obtained radial velocity measurements of TOI-849 with the HARPS spectrograph (R=115,000) mounted on the 3.6m telescope at ESO's La Silla Observatory \citep{2003Msngr.114...20M}. Thirty observations were taken between 28 July 2019 and 13 August 2019 in HAM mode, as part of the NCORES large programme (ID 1102.C-0249). An exposure time of 1200s was used for each observation, giving a signal-to-noise ratio of \mytilde20 per pixel. Typically the star was observed 2-3 times per night. The data were reduced with the offline DRS HARPS pipeline. RV measurements were derived using a weighted cross-correlation function (CCF) method using a G2V template \citep{1996A&AS..119..373B,2002Msngr.110....9P}, and the uncertainties in the RVs were estimated as described in \citet{2001A&A...374..733B}. The line bisector (BIS), and the full width half maximum (FWHM) were measured using the methods of \citet{2011A&A...528A...4B}. No correlation was seen between the RVs and calculated BIS, FWHM, or CCF contrast (R$<0.09$ in all cases). RV measurements are reported in Supplementary Table 1, and the RV data, photometry and best fit are shown in Figure \ref{figbestfit}. A jitter of $4.2ms^{-1}$ was seen, consistent with the low photometric activity level. BIS and FWHM are shown in Supplementary Figure 2. We investigated the CCFs for contributions from unresolved stellar companion by removing Gaussian fits to the individual CCF profiles and studying the residuals (Supplementary Figure 3). No evidence for additional companions is seen.

\subsection{LCOGT and PEST}

Two full transits of TOI-849 were observed on the nights UT 2019 July 30 and 2019 August 09 in $i'$ band using exposure times of 30 and 40 seconds, respectively. An additional night of data was taken on UT 2019 July 14, which unfortunately missed the transit relative to the revised ephemeris from our joint fit. The nights with transit are plotted in Supplementary Figure 1. Both observations used the CTIO node of the Las Cumbres Observatory Global Telescope (LCOGT) 1\,m network \citep{Brown:2013}. We used the {\tt TESS Transit Finder}, which is a customised version of the {\tt Tapir} software package \citep{Jensen:2013}, to schedule our transit observations. The telescopes are equipped with $4096\times4096$ LCO SINISTRO cameras having an image scale of $0\farcs389$ pixel$^{-1}$ resulting in a $26\arcmin\times26\arcmin$ field of view. The images were calibrated by the standard LCOGT BANZAI pipeline and the photometric data were extracted using the {\tt AstroImageJ} software package \citep{Collins:2017}. The first full transit on July 30 was observed with the telescope in focus and achieved a PSF FWHM of $\sim 1\farcs6$. Circular apertures with radius $3\farcs1$ were used to extract differential photometry for the target star and all stars within $2\farcm5$ that are brighter than \TESS band magnitude 19. All of the neighbouring stars were excluded as possible sources of the TESS detection, and the event was detected on target. A circular aperture with radius $8\arcsec$ was used for the other LCOGT observation, which was slightly defocused to a FWHM of $\sim4\arcsec$. The nearest star in the GAIA DR2 catalogue is $39\arcsec$ to the north of TOI-849, so the target star photometric apertures are uncontaminated by known nearby stars. 

A full transit was observed on UT 2019 August 20 in $\rm R_c$ band from the Perth Exoplanet Survey Telescope (PEST) near Perth, Australia. The 0.3 m telescope is equipped with a $1530\times1020$ SBIG ST-8XME camera with an image scale of 1$\farcs$2 pixel$^{-1}$, resulting in a $31\arcmin\times21\arcmin$ field of view. Systematics at the level of the shallow transit depth precluded inclusion of these data in the joint fit.

\subsection{NaCo/VLT}
TOI-849 was imaged with NaCo on the night of 2019 August 14 in NGS mode with the Ks filter. We took 9 frames with an integration time of 17s each, and dithered between each frame. We performed a standard reduction using a custom IDL pipeline: we subtracted flats and constructed a sky background from the dithered science frames, aligned and co-added the images, then injected fake companions to determine a 5$\sigma$ detection threshold as a function of radius. We obtained a contrast of 5.6 magnitudes at 1\arcsec, and no companions were detected. The contrast curve is shown in Supplementary Figure 4.

\subsection{SOAR}
We searched for nearby sources to TOI-849 with SOAR speckle imaging \citep{Tokovinin-18} on 12 August 2019 UT, observing in a similar visible bandpass as TESS. Additional details of the observation are available in \citet{Ziegler-19}. We detected no nearby sources within 3\arcsec of TOI-849. The $5\sigma$ detection sensitivity and the speckle auto-correlation function from the SOAR observation are plotted in Supplementary Figure 4.

\subsection{AstraLux}
We obtained a high-spatial resolution image of TOI-849 with the AstraLux camera \citep{Hormuth-08} installed at the 2.2m telescope of Calar Alto Observatory (Almería, Spain), using the lucky-imaging technique \citep{fried78}. We obtained 24\,400 images in the SDSSz band of 20 ms exposure time, well below the coherence time. The CCD was windowed to match 6$\times$6 \arcsec. We used the observatory pipeline to perform basic reduction of the images and subsequent selection of the best-quality frames. This is done by measuring their Strehl-ratio \citep{strehl1902} and selecting only the 10\% with the highest value of this parameter (thus an effective integration time of 48\,s). Then, these images are aligned and combined to obtain the final high-spatial resolution image. We estimate the sensitivity curve of this high-spatial resolution image by following the process explained in \citet{lillo-box12,lillo-box14b}, based on the injection of artificial stars in the image at different angular separations and position angles and measuring the retrieved stars based on the same detection algorithms used to look for real companions. No companions are detected in this image within the sensitivity limits. Both the high-resolution image and the contrast curve are shown in Supplementary Figure 4.

\subsection{Spectroscopic analysis and chemical abundances}     
\label{sec:parameters}

The spectroscopic analysis to derive the T$_{eff}$, $\log{g}$, microturbulence and [Fe/H]) and respective errors followed the methodology described in \citet[][]{Sousa-14, Santos-13}. Equivalent widths (EWs) are measured for a list of well defined iron lines. We used the combined HARPS spectrum of TOI-849 and ARES v2 code\footnote{The last version of ARES code (ARES v2) can be downloaded at http://www.astro.up.pt/$\sim$sousasag/ares} \citep{Sousa-07, Sousa-15} to perform the EW measurements. In the spectral analysis we look for the ionization and excitation equilibrium. The process makes use of a grid of Kurucz model atmospheres \citep{Kurucz-93} and the radiative transfer code MOOG \citep{Sneden-73}. The resulting values are T$_{eff}$= 5329$\pm$48, $\log{g}$= 4.28$\pm$0.09, $\xi_{t}$= 0.82$\pm$0.08, and [Fe/H]= +0.20$\pm$0.03.

The same tools and models were also used to derive stellar abundances for several chemical elements. For this we used the classical curve-of-growth analysis method assuming local thermodynamic equilibrium. Although the EWs of the spectral lines were automatically measured with ARES, for the elements with only two to three lines available we performed careful visual inspection of the EW measurements. For the derivation of chemical abundances we closely followed the methods described in \citet[e.g.][]{ Adibekyan-15}. The final abundances derived are [NaI/H]= 0.30$\pm$0.16, [MgI/H]= 0.24$\pm$0.06, [AlI/H]= 0.30$\pm$0.06, [SiI/H]= 0.24$\pm$0.08, [CaI/H]= 0.16$\pm$0.07, [ScII/H]= 0.23$\pm$0.09, [TiI/H]= 0.25$\pm$0.09, [CrI/H]= 0.23$\pm$0.07, and [NiI/H]= 0.28$\pm$0.04.

Supplementary Figure 5 shows a comparison of the abundances of TOI-849 with the ones found in the solar neighbourhood stars \citep[][]{Adibekyan-12} of similar atmospheric parameters. In terms of chemical composition TOI-849 seems to be very similar to the solar neighbourhood stars showing slight enhancement in the iron-peak elements Cr and Ni.

\subsection{Joint RV and photometric fit}

The HARPS RVs, the \textit{TESS}, NGTS and LCOGT photometry and the spectral energy distribution (SED) were jointly analysed in a Bayesian framework, using the \texttt{PASTIS} software \citep{Diaz:2014kd,Santerne:2015bb}. For the SED, we used the visible magnitudes from the American Association of Variable Star Observers Photometric All-Sky Survey (APASS) and the near-infrared magnitudes from the Two-Micron All-Sky Survey (2MASS) and the Wide-field Infrared Survey Explorer (AllWISE) \citep{2015AAS...22533616H, 2014AJ....148...81M, 2014yCat.2328....0C}. The RVs were fitted using a Keplerian orbit model and a linear drift. The light curves were modelled with the JKT Eclipsing Binary Orbit Program \citep[JKTEBOP,][]{2008MNRAS.386.1644S} using an oversampling factor of 180, 12, 6, and 7 for the TESS and the three LCOGT-CTIO light curves, respectively. The NGTS light curves were not oversampled as the integration of the individual data is short with respect to the transit duration \citep{2010MNRAS.408.1758K}. Finally, the SED was modelled with the BT-Settl library of stellar atmosphere models \citep{2012RSPTA.370.2765A}. The system parameters and associated uncertainties were derived using the Markov Chain Monte Carlo (MCMC) method implemented in \texttt{PASTIS}. The stellar parameters were computed using the Dartmouth evolution tracks \citep{2008ApJS..178...89D} at each step of the chains, accounting for the asterodensity profiling \citep{2014MNRAS.440.2164K}. We also used the PARSEC evolution tracks, with consistent results.

Regarding the priors, we used a Normal distribution with median and width from the spectral analysis for the stellar temperature, surface gravity and iron abundance. For the systemic distance to Earth, we used a normal prior centered on the \textit{Gaia} Data Release 2 \citep{2018A&A...616A...1G} value, taking into account the correction from \citet{2019MNRAS.487.3568S}. For the orbital period and transit epoch, we used Normal priors centered on first guess values from an independent analysis of the NGTS and \TESS light curves alone, to improve the convergence of the MCMCs. For the orbital inclination, we used a sine prior and for the eccentricity a truncated normal prior with width 0.083, following \citep{2019AJ....157...61V}. For the other parameters, we used uniform priors with width large enough to not artificially decrease the uncertainties. Initial fits gave an insignificant eccentricity of $0.033^{+0.025}_{-0.021}$ and so we fixed eccentricity to zero for final fitting. A marginally significant linear drift was included for the HARPS data, and did not affect the results.

We ran 20 MCMCs with $2 \times 10^5$ iterations. We checked the convergence with a Kolmogorov-Smirnov test \citep{Diaz:2014kd,Santerne:2015bb}, removed the burn-in phase and merged the remaining chains. The limb darkening coefficients were computed using the stellar parameters and tables from \citet{2011A&A...529A..75C}. Finally, the physical parameters and associated uncertainties were derived from samples from the merged chain. The results for the Dartmouth and PARSEC evolution tracks can be seen in Table \ref{MCMCprior}.

As an independent check on the derived stellar parameters, we performed an analysis of the broadband spectral energy distribution (SED) together with the {\it Gaia\/} parallax in order to determine an empirical measurement of the stellar radius, following the procedures described in \citet{Stassun-16,Stassun-17,Stassun-18}. We pulled the $B_T V_T$ magnitudes from {\it Tycho-2}, the $BVgri$ magnitudes from APASS, the $JHK_S$ magnitudes from {\it 2MASS}, the W1--W4 magnitudes from {\it WISE}, and the $G$ magnitude from {\it Gaia}. Together, the available photometry spans the full stellar SED over the wavelength range 0.4--22~$\mu$m. We also checked the {\it GALEX} NUV flux, which was not used in the fit as it suggests a modest level of chromospheric activity. 

We performed the independent fit using the Kurucz stellar atmosphere models, with the priors on effective temperature ($T_{\rm eff}$), surface gravity ($\log g$), and metallicity ([Fe/H]) from the spectroscopic values. The remaining free parameter is the extinction ($A_V$), which we limited to the maximum line-of-sight extinction from the \citet{Schlegel-98} dust maps. The resulting fit has a reduced $\chi^2$ of 4.5, and a best fit extinction of $A_V = 0.04 \pm 0.03$. Integrating the (unextincted) model SED gives the bolometric flux at Earth of $F_{\rm bol} = 3.713 \pm 0.086 \times 10^{-10}$ erg~s~cm$^{-2}$. Taking the $F_{\rm bol}$ and $T_{\rm eff}$ together with the {\it Gaia\/} parallax, adjusted by $+0.08$~mas to account for the systematic offset reported by \citet{StassunTorres-18}, gives the stellar radius as $R = 0.896 \pm 0.020$~R$_\odot$. Finally, estimating the stellar mass from the empirical relations of \citet{Torres-10}, assuming solar metallicity, gives $M = 1.01 \pm 0.08 M_\odot$, which with the radius gives the mean stellar density $\rho = 1.99 \pm 0.19$ g~cm$^{-3}$. These values are consistent with the stellar parameters found as part of the \texttt{PASTIS} MCMC chain, and so we adopt the \texttt{PASTIS} values for our results.

\subsection{Interior structure characterisation}

Given the mass and radius of TOI-849b it is clear that the planet does not represent a larger version of Neptune. This is demonstrated in Figure \ref{figMRplot} which shows the M-R relation for a pure-water curve and a planet with 95\% water and 5\% H-He atmosphere corresponding to a stellar irradiation of $F/F_{\oplus}=3000$ (TOI-849b). TOI-849b sits on the pure-water curve and well below the 5\% strongly irradiated curve, suggesting that the H-He mass fraction is of the order of only a few percent, if not negligible. Figure \ref{figNepDesert} also shows that TOI-849b is relatively isolated in parameter space, suggesting that it is somewhat unique and could have been subjected to an unusually aggressive removal of the primordial H-He envelope.

We explore layered structure models containing variable fractions of H-He envelope. Typical available models are not suited to this planet due to the high pressures in the interior, requiring exotic equations of state. Further, for planets this massive the interior layers are likely not distinct as for smaller planets, with composition gradients more likely \citep{Bodenheimer-18}. Rather than build a full model of the interior, which would not be valid for the reasons stated, we consider some illuminating limiting cases. 

We model the planetary interior of TOI-849b assuming a pure iron core, a silicate mantle, a pure water layer, and a H-He atmosphere. We follow the structure model of \citet{Dorn-17A} except for the iron core, for which we use the updated EOS presented in \citet{Hakim-18A}. For the silicate-mantle, equilibrium mineralogy and density are computed as a function of pressure, temperature, and bulk composition by minimizing Gibbs free energy \cite[e.g.][]{Connolly09}. For the water we use the quotidian equation of state (QEOS) presented in \citet{Vazan13} for low pressures and the tabulated EoS from \citet{Seager07} for pressures above 44.3 GPa. For H-He we use SCVH \citep{Saumon95} assuming a proto-solar composition. We then solve the standard structure equations. 

We then estimate the possible range of H-He mass fraction in TOI-849b which fits the derived mass and radius. In order to estimate the maximum possible mass of an H-He envelope, we assume a planet without water. The core-to-mantle fraction is set by the stellar abundance [Fe/Si] of the host star \citep{Dorn15}.  
The minimum H-He mass fraction is estimated by assuming a large fraction of water of 70\% by mass, which corresponds to a  water-rich planet. We search for the maximum and minimum H-He mass fractions for a grid of planetary masses and radii covering the observed values and their 2-$\sigma$ error range.
It is found that that H-He mass fraction is at minimum $2.8^{+0.8} _{-1.0}$\% and at maximum $3.9^{+0.8} _{-0.9}$\%, suggesting that the heavy-element mass is above $39 M_{\oplus}$. 
It should be noted that our models assume a pure H-He atmosphere, while in reality the atmosphere is expected to include heavier elements as inferred by recent formation models \citep[e.g.][]{Lozovsky17,Bodenheimer-18}. This is particularly true for planets this massive where the interior layers are likely not distinct as for smaller planets. The existence of heavy elements in the H-He atmosphere would lead to compression, and can therefore increase the planetary H-He mass fraction. However, for the case of TOI-849b, the difference is expected to be very moderate since the planet mass is clearly dominated by heavy-elements. \citet{Lovozsky18} calculated the effect of varying atmospheric water content on planetary radii for fixed masses and H-He gas mass fractions. Applying their model to TOI-849b showed that the inferred planet radius is only affected on the few percent level for atmospheric water content ranging from 0 to 70\%. As such we expect the plausible increase in H-He to be small even for high levels of volatile enrichment in the planetary envelope. We can therefore conclude that the mass fraction of H-He is at most a few percent.

\subsection{Photoevaporation Rate}

We explored the X-ray and EUV irradiation of the planet, wavelengths most relevant for atmospheric mass loss \cite[e.g.][]{Lammer-03}. Archival X-ray data exists for the system only from the \textit{ROSAT} All-Sky Survey, where the nearest detected source is an arcminute away, too far to be associated with TOI-849. Instead, we applied the empirical relations of \citet{Jackson-12} linking X-ray emission with age, estimating $L_{\rm X}/L_{\rm bol} = 7.5\times10^{-7}$ at the current age. This figure implies an X-ray flux at Earth of $3.0\times10^{-16}$\,erg\,s$^{-1}$\,cm$^{-2}$, much too faint to be visible with \textit{XMM-Newton} or \textit{Chandra}. We extrapolated our X-ray estimate to the unobservable EUV band using the relations of \citet{King-18}, based on the method of \citet{Chadney-15}.

To estimate mass loss rates, we applied both the energy-limited approach \citep{Watson-81,Erkaev-07}, and a method based on interpolating and approximating to hydrodynamical simulations \citep{Kubyshkina-18A,Kubyshkina-18B}. The latter yields a loss rate of $1.8\times10^{11}$\,g\,s$^{-1}$, more than an order of magnitude larger than the former when assuming a canonical efficiency of 15\%. Integrating over the planet's XUV history, and starting at a Jupiter mass and radius, we estimate total lifetime losses of 4.0\% and 0.81\% of the planet's mass using the energy-limited and Kubyshkina methods, respectively. While these calculations have the limitation of assuming a constant radius across the lifetime, these losses are not enough to evolve the planet to one slightly smaller than Neptune, and so we can be sure the planet did not start as a Jupiter-like giant if its evolution has been solely through photoevaporation.

An intermediate starting point is the planet HD149026b \citep{Sato-05}, a giant planet with mass $121\pm19M_\oplus$ and radius $8.3\pm0.2R_\oplus$ \citep{Stassun-17}. For this planet, we estimate total lifetime losses of 11.42\% and 100\% of the planet's mass using the energy-limited and Kubyshkina methods, respectively. These are likely to be significant overestimates, due to the constant radius assumption which clearly becomes flawed after significant mass loss. As such finding the limits of photoevaporation in creating a planet like TOI-849b requires detailed models beyond the scope of this paper.

\subsection{Planet Population Synthesis Models}

We explored possible formation channels for such dense Neptune sized planets via the Bern planetary population synthesis models \citep{2018haex.bookE.143M}. An updated form of the models was used, with particularly relevant changes being:
\begin{enumerate}
\item An improved gas disk surface temperature from the stellar evolution tracks of \citet{Baraffe-15}, with a vertically-integrated approach to compute the vertical structure of the gas disc \citep{Nakamoto-94}.
\item A higher concentration of solids in the inner part of the disc following \citet{Ansdell-18}
\item Gas-driven type I migration computed following \citet{Coleman-14}
\item Giant impacts induce an additional luminosity similar to \citet{Broeg-12}
\end{enumerate}

In those models, which were run before the discovery of TOI-849b, we found four planets that exhibit similar mass, radius and eccentricity to TOI-849b, out of a total sample of 1000. These planets have masses between 20 and 50$M_\oplus$ and have an ice content of 20-30\% by mass, but no H/He. They started as embryos outside the ice line, and migrated steadily to a position close to the inner edge of the disc. The removal of the primordial H/He is due to a giant impact that takes place at the end of the migration, which means that the planets are unable to accrete a second H/He envelope. Due the high equilibrium temperature, it is likely that the ices evaporate to form a secondary atmosphere consisting of water and possibly other volatiles like CO and CO$_2$. Such an envelope leads to radii comparable to the discovered planet. From the modelling point of view, the population synthesis models thus prefer planets whose small envelopes consist entirely of ices. The evolution tracks of the four considered model planets are shown in Supplementary Figure 6.

Although no similar model planets to TOI-849b were found from other formation pathways, this should not be taken as evidence against other hypotheses such as gap opening limiting the accretion, or tidal disruption. The Bern models do not include gap opening in the disk as a limiting factor in gas accretion, and use simplified assumptions for tidal interactions following \citet{Benitez-11} that do not include high eccentricity migration.

\subsection{Tidally induced thermalisation events}

The high bulk density of TOI-849b (5.5 g/cm$^3$) relative to Neptune (1.6 g/cm$^3$) suggests that the planet (with a radius equal to 90\% of Neptune's) might currently represent the core of a previously giant planet. For this scenario to be viable, the planet needed to originate as a gas giant and have expelled mass, possibly during orbit shrinkage and circularization. This evolutionary pathway may occur as a result of chaotic tides \citep{2004MNRAS.347..437I,2018MNRAS.476..482V,2018AJ....155..118W}, where the planet's internal f-modes were excited after the planet was gravitationally scattered onto a highly eccentric orbit. Energy build up in the modes could have then led to thermalisation events, potentially ejecting atmospheric layers \citep{2019MNRAS.484.5645V,2019MNRAS.489.2941V}. After the resulting core left the chaotic regime, subsequent orbital evolution over the $\sim 9$ Gyr main-sequence lifetime of the parent star may have proceeded with weakly dissipative equilibrium tides, leading to the current orbit. In this scenario, the planet may have expelled 1-2 orders of magnitude more mass than its current value.

Accumulation of the internal mode energy leads to thermalisation events, which subsequently deposits energy into the planet's interior and resets the mode amplitude. Possible results of the thermalisation events include inflation, mass ejection or both; TOI-849b could have experienced these events and still retained some or all of its atmosphere. Although the trigger for and consequences of these events remains largely unknown, \citet{2019MNRAS.484.5645V} assumed these events occur when the accumulated mode energy equals 10\% of the planet's binding energy

\begin{equation}
E_{\rm bind} \approx \frac{G M_{\rm p}^2}{R_{\rm p}}
.
\end{equation}

\noindent{}They also demonstrated that the resulting changes in orbital evolution due to the thermalisation events is largely independent of this choice of 10\%. With this choice, \citet{2019MNRAS.489.2941V} illustrated that the number of thermalisation events which a planet experiences is positively correlated with increasing puffiness of the planet and decreasing orbital pericentre. They showed that even a dense gas giant with a pericentre of about 1.5 Solar radii would experience at least one thermalisation event, albeit with a smaller mass central star. TOI-849b, which currently resides at a distance of about 3 Solar radii, previously would have harboured a pericentre that is just half of that value if angular momentum was conserved as its eccentricity decreased from almost unity to zero, under the high-eccentricity circularisation scenario. 

\newpage
\begin{longtable}{lllr}
\caption{\label{MCMCprior} List of parameters used in the analysis. The respective priors are provided together with the posteriors for the Dartmouth and PARSEC stellar evolution tracks. The posterior values represent the median and 68.3\% credible interval. Derived values that might be useful for follow-up work are also reported.}\\
\hline
Parameter & Prior & \multicolumn{2}{c}{Posterior}\\
 &  & Dartmouth & PARSEC \\
 &  & (adopted) & \\
\hline
\endfirsthead
\multicolumn{4}{l}{{\bfseries \tablename\ \thetable{} -- continued from previous page}} \\
\hline
Parameter & Prior & \multicolumn{2}{c}{Posterior}\\
 &  & Dartmouth & PARSEC\\
\hline
\endhead
\multicolumn{4}{l}{{Continued on next page}} \\ 
\hline
\endfoot
\hline
\multicolumn{4}{l}{Notes:}\\
\multicolumn{4}{l}{$\bullet$ $\mathcal{N}(\mu,\sigma^{2})$: Normal distribution with mean $\mu$ and width $\sigma^{2}$}\\
\multicolumn{4}{l}{$\bullet$ $\mathcal{U}(a,b)$: Uniform distribution between $a$ and $b$}\\
\multicolumn{4}{l}{$\bullet$ $\mathcal{S}(a,b)$: Sine distribution between $a$ and $b$}\\
\multicolumn{4}{l}{$\bullet$ $\mathcal{T}(\mu,\sigma^{2},a,b)$: Truncated normal distribution with mean $\mu$ and width $\sigma^{2}$, between $a$ and $b$}\\
\endlastfoot
\\
\multicolumn{4}{l}{\it Stellar Parameters}\\
\\
Effective temperature \teff\ [K] & $\mathcal{N}(5329.0, 48.0)$ & $5375.3^{_{+41.8}}_{^{-41.4}}$ & $5379.1^{_{+41.4}}_{^{-43.4}}$ \\
Surface gravity log~{\it g}\ [cgs] & $\mathcal{N}(4.43,0.3)$ & $4.48^{_{+0.03}}_{^{-0.04}}$ & $4.47^{_{+0.03}}_{^{-0.04}}$ \\
Iron abundance [Fe/H]\ [dex] &  $\mathcal{N}(0.201,0.033)$ & $0.19\pm0.03$ & $0.19\pm0.03$ \\
Distance to Earth $D$ [pc] &  $\mathcal{N}(224.56,7.1)$ & $225.2^{_{+6.1}}_{^{-5.8}}$ & $224.7^{_{+6.6}}_{^{-5.9}}$ \\
Interstellar extinction $E(B-V)$ [mag] &  $\mathcal{U}(0.0,1.0)$ & $0.011^{_{+0.016}}_{^{-0.009}}$ & $0.011^{_{+0.017}}_{^{-0.008}}$ \\
Systemic radial velocity $\gamma$ [$kms^{-1}s^{-1}$] & $\mathcal{U}(5.0,15.0)$ & $9.3503\pm0.0012$ & $9.3503\pm0.0012$ \\
Stellar density $\rho_{\star}/\rho_{\odot}$ & (derived) & $1.197^{_{+0.109}}_{^{-0.132}}$ & $1.179^{_{+0.120}}_{^{-0.145}}$ \\
Stellar mass M$_{\star}$\ [$M_\odot$] & (derived) & $0.929^{_{+0.023}}_{^{-0.023}}$ & $0.903^{_{+0.027}}_{^{-0.029}}$ \\
Stellar radius R$_{\star}$\ [$R_\odot$] & (derived) & $0.919^{_{+0.031}}_{^{-0.022}}$ & $0.915^{_{+0.033}}_{^{-0.022}}$ \\
Stellar age $\tau$\ [Gyr] & (derived) & $6.7^{_{+2.8}}_{^{-2.4}}$ & $8.5^{_{+3.6}}_{^{-3.0}}$ \\
&&& \\
\hline
\\
\multicolumn{4}{l}{\it Planet b Parameters}\\
\\
Orbital Period $P_{b}$ [d] &  $\mathcal{N}(0.76552484,0.00000435)$ & $0.76552398^{_{+0.00000262}}_{^{-0.00000273}}$ & $0.76552403^{_{+0.00000260}}_{^{-0.00000278}}$ \\
Epoch $T_{0,b}$ [BJD - 2450000] &  $\mathcal{N}(8394.73741796,0.0017159129)$ & $8394.73767^{_{+0.00096}}_{^{-0.00095}}$ & $8394.73765^{_{+0.00093}}_{^{-0.00093}}$ \\
RV semi-amplitude $K_{b}$ [km s$^{-1}$] & $\mathcal{U}(0.0,0.1)$ & $0.02988^{_{+0.00167}}_{^{-0.00173}}$ & $0.02992^{_{+0.00170}}_{^{-0.00176}}$ \\
Orbital inclination $i_{b}$ [$^{\circ}$] & $\mathcal{S}(50.0,90.0)$ & $86.8^{_{+2.2}}_{^{-2.6}}$ & $86.5^{_{+2.4}}_{^{-2.8}}$ \\
Planet-to-star radius ratio $k_{b}$ & $\mathcal{U}(0.0,1.0)$ & $0.03444^{_{+0.00091}}_{^{-0.00092}}$ & $0.03444^{_{+0.00095}}_{^{-0.00090}}$ \\
Orbital eccentricity $e_{b}$ & $\mathcal{T}(0.0,0.083,0.0,1.0)$ & $0.0\pm0.0$ & $0.0\pm0.0$ \\
Argument of periastron $\omega_{b}$ [\degr] & $\mathcal{U}(0.0,360.0)$ & $0.0\pm0.0$ & $0.0\pm0.0$ \\
System scale $a_{b}/R_{\star}$ & (derived) & $3.7^{_{+0.1}}_{^{-0.1}}$ & $3.7^{_{+0.1}}_{^{-0.2}}$ \\
Impact parameter $b_{b}$ & (derived) & $0.212^{_{+0.158}}_{^{-0.143}}$ & $0.228^{_{+0.166}}_{^{-0.152}}$ \\
Transit duration T$_{14,b}$ [h] & (derived) & $1.57\pm0.04$ & $1.57\pm0.04$ \\
Semi-major axis $a_{b}$ [AU] & (derived) & $0.01598^{_{+0.00013}}_{^{-0.00013}}$ & $0.01583^{_{+0.00016}}_{^{-0.00017}}$ \\
Planet mass M$_{b}$ [$M_\oplus$] & (derived) & $40.78^{_{+2.41}}_{^{-2.45}}$ & $40.03^{_{+2.48}}_{^{-2.41}}$ \\
Planet radius R$_{b}$ [$R_\oplus$] & (derived) & $3.447^{_{+0.164}}_{^{-0.122}}$ & $3.432^{_{+0.177}}_{^{-0.127}}$ \\
Planet bulk density $\rho_{b}$ [g cm$^{-3}$] & (derived) & $5.5^{_{+0.8}}_{^{-0.8}}$ & $5.4^{_{+0.8}}_{^{-0.9}}$ \\
&&& \\
\hline
&&& \\
\multicolumn{4}{l}{\it Instrument-related Parameters}\\
&&& \\
HARPS jitter [$kms^{-1}s^{-1}$] & $\mathcal{U}(0.0,0.1)$ & $0.00422^{_{+0.00127}}_{^{-0.00118}}$ & $0.00425^{_{+0.00134}}_{^{-0.00118}}$ \\
HARPS drift [$kms^{-1}d^{-1}$] & $\mathcal{U}(-0.001,0.001)$ & $0.00054\pm0.00022$ & $0.00054\pm0.00023$ \\
\textit{TESS} contamination [\%] & $\mathcal{T}(0.0,0.005,0.0,1.0)$ & $0.003^{_{+0.004}}_{^{-0.002}}$ & $0.003^{_{+0.004}}_{^{-0.002}}$ \\
\textit{TESS} jitter [ppm] & $\mathcal{U}(0.0, 10^5)$ & $54.1^{_{+53.3}}_{^{-37.4}}$ & $52.2^{_{+54.4}}_{^{-36.5}}$ \\
\textit{TESS} out-of-transit flux & $\mathcal{U}(0.99,1.01)$ & $1.0001002^{_{+0.0000225}}_{^{-0.0000217}}$ & $1.0001003^{_{+0.0000218}}_{^{-0.0000218}}$ \\
\textit{TESS} limb-darkening $u_{a}$ & (derived) & $0.3764\pm0.0072$ & $0.3756\pm0.0074$ \\
\textit{TESS} limb-darkening $u_{b}$ & (derived) & $0.2387\pm0.0041$ & $0.2391\pm0.0042$ \\
\textit{NGTS}$_1$ contamination [\%] & $\mathcal{T}(0.0,0.005,0.0,1.0)$ & $0.003^{_{+0.004}}_{^{-0.002}}$ & $0.003^{_{+0.004}}_{^{-0.002}}$ \\
\textit{NGTS}$_1$ jitter [ppm] & $\mathcal{U}(0.0, 10^5)$ & $78.5^{_{+87.1}}_{^{-55.9}}$ & $78.4^{_{+82.8}}_{^{-55.4}}$ \\
\textit{NGTS}$_1$ out-of-transit flux & $\mathcal{U}(0.99,1.01)$ & $1.0000812^{_{+0.0000838}}_{^{-0.0000878}}$ & $1.0000794^{_{+0.0000888}}_{^{-0.0000843}}$ \\
\textit{NGTS}$_2$ contamination [\%] & $\mathcal{T}(0.0,0.005,0.0,1.0)$ & $0.003^{_{+0.004}}_{^{-0.002}}$ & $0.003^{_{+0.004}}_{^{-0.002}}$ \\
\textit{NGTS}$_2$ jitter [ppm] & $\mathcal{U}(0.0, 10^5)$ & $86.6^{_{+93.6}}_{^{-61.4}}$ & $84.9^{_{+92.6}}_{^{-60.3}}$ \\
\textit{NGTS}$_2$ out-of-transit flux & $\mathcal{U}(0.99,1.01)$ & $1.0000772^{_{+0.0000987}}_{^{-0.0000958}}$ & $1.0000742^{_{+0.0000959}}_{^{-0.0000938}}$ \\
\textit{NGTS} limb-darkening $u_{a}$ & (derived) & $0.4755\pm0.0081$ & $0.4748\pm0.0084$ \\
\textit{NGTS} limb-darkening $u_{b}$ & (derived) & $0.2116\pm0.0051$ & $0.2121\pm0.0052$ \\
\textit{LCO}$_1$ contamination [\%] & $\mathcal{T}(0.0,0.005,0.0,1.0)$ & $0.003^{_{+0.004}}_{^{-0.002}}$ & $0.003^{_{+0.004}}_{^{-0.002}}$ \\
\textit{LCO}$_1$ jitter [ppm] & $\mathcal{U}(0.0, 10^5)$ & $1021.9^{_{+90.7}}_{^{-89.0}}$ & $1020.0^{_{+89.9}}_{^{-84.6}}$ \\
\textit{LCO}$_1$ out-of-transit flux & $\mathcal{U}(0.98,1.02)$ & $0.9999932^{_{+0.0000854}}_{^{-0.0000851}}$ & $0.9999967^{_{+0.0000891}}_{^{-0.0000916}}$ \\
\textit{LCO}$_2$ contamination [\%] & $\mathcal{T}(0.0,0.005,0.0,1.0)$ & $0.003^{_{+0.004}}_{^{-0.002}}$ & $0.003^{_{+0.004}}_{^{-0.002}}$ \\
\textit{LCO}$_2$ jitter [ppm] & $\mathcal{U}(0.0, 10^5)$ & $1421.7^{_{+84.0}}_{^{-84.8}}$ & $1419.3^{_{+84.0}}_{^{-82.4}}$ \\
\textit{LCO}$_2$ out-of-transit flux & $\mathcal{U}(0.98,1.02)$ & $0.9999893^{_{+0.0001046}}_{^{-0.0001040}}$ & $0.9999905^{_{+0.0001043}}_{^{-0.0001014}}$ \\
\textit{LCO} limb-darkening $u_{a}$ & (derived) & $0.3826\pm0.0074$ & $0.3818\pm0.0076$ \\
\textit{LCO} limb-darkening $u_{b}$ & (derived) & $0.2388\pm0.0043$ & $0.2392\pm0.0042$ \\
SED jitter [mag]  & $\mathcal{U}(0.0,0.1)$ & $0.047^{_{+0.03}}_{^{-0.026}}$ & $0.047^{_{+0.03}}_{^{-0.026}}$ \\
&&& \\
\end{longtable}

\newpage

\begin{table}
\caption{Stellar Properties of TOI-849}
\label{tabstellarproperties}
\begin{threeparttable}
\begin{tabular}{llr}
\hline
Property & Value & Source \\
\hline
\multicolumn{2}{l}{\textbf{Astrometric Properties}}  \\
RA & 01:54:51.7910 & GAIA DR2\tnote{1}\\
Dec & -29:25:18.1508 & GAIA DR2\tnote{1}\\
TIC ID & 33595516 & TICv8\tnote{2}\\
GAIA ID & 5023809953208388352& GAIA DR2\tnote{1}\\
2MASS ID & 01545169-2925186 & 2MASS\tnote{3}\\
$\mu_\textrm{RA}$ (mas.yr$^{-1}$)   & 73.315 & GAIA DR2\tnote{1}\\
$\mu_\textrm{Dec}$ (mas.yr$^{-1}$)   & 20.664 & GAIA DR2\tnote{1}\\
\hline
\\
\multicolumn{2}{l}{\textbf{Photometric Properties}} \\
TESS (mag) & 11.55 & TICv8\tnote{2}\\
B (mag) & 12.84 & TICv8\tnote{2}\\
V (mag) & 11.98 &TICv8\tnote{2}\\
G (mag) & 12.06 &TICv8\tnote{2}\\
J (mag) & 10.83 & TICv8\tnote{2}\\
H (mag) & 10.48 &TICv8\tnote{2}\\
K (mag) & 10.42 & TICv8\tnote{2}\\
\hline
\end{tabular}
\begin{tablenotes}[flushleft]\footnotesize
    \item[1] \citet{2018A&A...616A...1G}
    \item[2] \citet{2019AJ....158..138S}
    \item[3] \citet{2006AJ....131.1163S}
\end{tablenotes}
\end{threeparttable}
\end{table}

\bibliography{papers031019_plus}
\bibliographystyle{mn2e_fix}

\section{addendum}

\subsection{Author Contributions}
DJArm is PI of the NCORES HARPS programme which measured the planet's mass, a member of the NGTS consortium, developed much of the text and main figures and coordinated all contributions.
TLop performed the joint PASTIS analysis.
VAdi, SSou, NSan performed stellar spectral analysis including chemical abundances.
RBoo, FMer provided text analysing potential formation scenarios.
KACol, EJen coordinated the TFOP SG1 photometric followup of the system.
KICol, TGan, performed analysis of LCOGT photometric followup of the system. 
AEms, CMor performed and analyses the Bern Population Synthesis Models.
CHua, LSha developed and ran the MIT Quick Look Pipeline which identified the candidate in the TESS data.
GKin performed the photoevaporation analysis.
JLil obtained and analyses the Astralux data, and synthesised all HR imaging results.
EMat obtained the NaCo imaging data.
HOsb contributed to the NCORES HARPS programme and the NGTS survey, and contributed to the main figures.
JOte, OMou, MDel, RHel, MLoz, CDor performed interior structure calculations.
DVer performed analysis on the potential for tidal self-disruption.
CZie obtained the SOAR data and provided text summarising SOAR results.
TGTan obtained a further transit with the PEST telescope.
JLiss contributed to the internal structure discussion.
KSta provided the independent check of stellar parameters.
MBro, SGan calculated estimates of required telescope time for atmospheric characterisation.
DRAnd, MMoy contributed to the maintenance and operation of the NGTS facility.
EBry, CWat, JSJen, JIVin, JAct, DBay, CBel, MBur, SCas, ACha, PEig, SGil, MGoa, MGue, MLen, JMcC, DPol, DQue, LRay, RTil, RWes contributed to the NGTS facility, either in planning, management, data collection or detrending.
DJABro, SHoj, DBar, SCCBar, PAW, LNie, DBay, FBou, BCoo, RDia, ODem, XDum, PFig, JJac, GKen, ASan, SUdr, PWil, JAlm contributed to the HARPS large programme under which HARPS data was obtained.
DCia, ICro, JSch, SHow contributed to the NaCo imaging data.
CBri, NLaw, AMan contributed to the SOAR imaging data.
KDCol, MFau, JoJen, EJen, GRic, PRow, SSea, ETin, RVan, JWin, JNVil, ZZan provided essential contributions to the \TESS mission which discovered the candidate.
All authors read the manuscript and provided general comments.

 \subsection{Acknowledgments} 
This paper includes data collected by the TESS missions, which are publicly available from the Mikulski Archive for Space Telescopes (MAST). Funding for the TESS mission is provided by NASA\'s Science Mission directorate. We acknowledge the use of public TESS Alert data from pipelines at the TESS Science Office and at the TESS Science Processing Operations Center. Resources supporting this work were provided by the NASA High-End Computing (HEC) Program through the NASA Advanced Supercomputing (NAS) Division at Ames Research Center for the production of the SPOC data products. This research has made use of the Exoplanet Follow-up Observation Program website and the NASA Exoplanet Archive, which are operated by the California Institute of Technology, under contract with the National Aeronautics and Space Administration under the Exoplanet Exploration Program. This work makes use of observations from the LCOGT network. Based on observations made with ESO Telescopes at the La Silla Paranal Observatory under programme ID 1102.C-0249. Based in part on observations collected at the European Organisation for Astronomical Research in the Southern Hemisphere under ESO program P103.C-0449. DJA, DV and SLC respectively acknowledge support from the STFC via Ernest Rutherford Fellowships (ST/R00384X/1), (ST/P003850/1) and (ST/R003726/1). GMK is supported by the Royal Society as a Royal Society University Research Fellow. FM. acknowledges support from the Royal Society Dorothy Hodgkin Fellowship. K.G.S.\ acknowledges partial support from NASA grant 17-XRP17 2-0024. C.Z. is supported by a Dunlap Fellowship at the Dunlap Institute for Astronomy \& Astrophysics, funded through an endowment established by the Dunlap family and the University of Toronto. A.W.M was supported by NASA grant 80NSSC19K0097 to the University of North Carolina at Chapel Hill. DJAB acknowledges support from the UK Space Agency. CXH and MNG acknowledge support from the Juan Carlos Torres Fellowship. This work was financed by FEDER - Fundo Europeu de Desenvolvimento Regional funds through the COMPETE 2020 - Operacional Programme for Competitiveness and Internationalisation (POCI), and by Portuguese funds through FCT - Funda\c{c}\~ao para a Ci\^encia e a Tecnologia in the framework of the projects 
UID/FIS/04434/2019; PTDC/FIS-AST/32113/2017 \& POCI-01-0145-FEDER-032113; PTDC/FIS-AST/28953/2017 \& POCI-01-0145-FEDER-028953. SSou, VAdi, SCCBar, ODSD acknowledge support from FCT through Investigador FCT contracts nr. IF/00028/2014/CP1215/CT0002, IF/00650/2015/CP1273/CT0001, and IF/01312/2014/CP1215/CT0004, DL 57/2016/CP1364/CT0004. SHoj acknowledge support by the fellowships PD/BD/128119/2016 funded by FCT (Portugal). Work by JNW was partly funded by the Heising-Simons Foundation. CAW would like to acknowledge support from UK Science Technology and Facility Council grant ST/P000312/1. JLil and DBar are funded by the Spanish State Research Agency (AEI) Projects No.ESP2017-87676-C5-1-R and No. MDM-2017-0737 Unidad de Excelencia “Mar\'ia de Maeztu”- Centro de Astrobiolog\'ia (INTA-CSIC). JSJ acknowledges funding by Fondecyt through grant 1161218 and partial support from CATA-Basal (PB06, Conicyt). JIV acknowledges support of CONICYT-PFCHA/Doctorado Nacional-21191829, Chile. The French group acknowledges financial support from the French Programme National de Plan\'etologie (PNP, INSU).  FM acknowledges support from the Royal Society Dorothy Hodgkin Fellowship.
 
 \subsection{Competing Interests} The authors declare that they have no
competing financial interests.

 \subsection{Correspondence} Correspondence and requests for materials
should be addressed to D.J.Armstrong.~(email: d.j.armstrong@warwick.ac.uk).

\section{Extended Data}

\renewcommand{\figurename}{Extended Data Figure}
\setcounter{figure}{0}

\begin{figure}
\resizebox{\hsize}{!}{\includegraphics{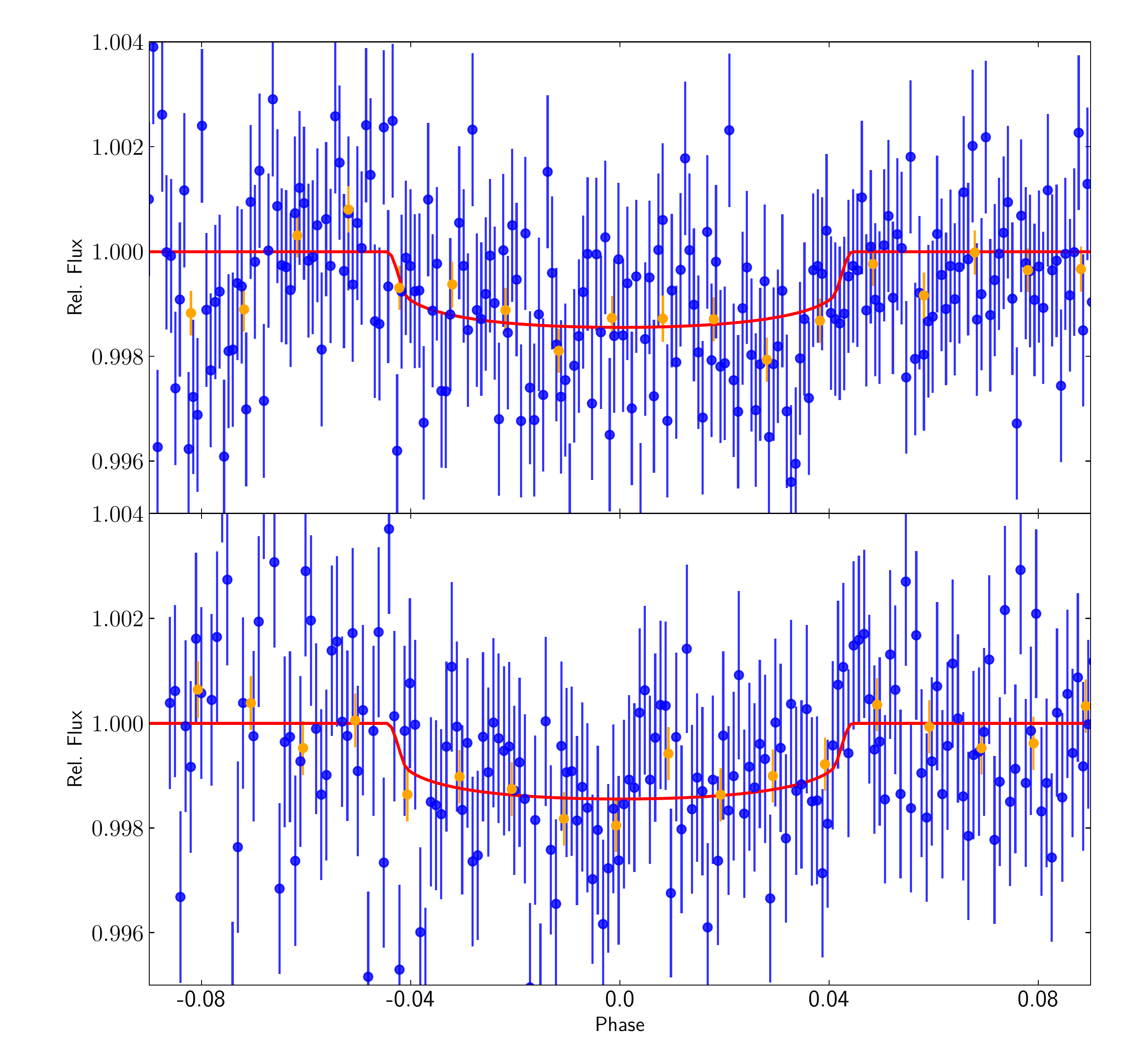}}
\caption{Photometric data captured by the LCOGT network on the nights UT 2019 July 30 (top) and 2019 August 09 (bottom). The best fit model is plotted in red and binned data in orange.}
\label{}
\end{figure}

\begin{figure}
\resizebox{\hsize}{!}{\includegraphics{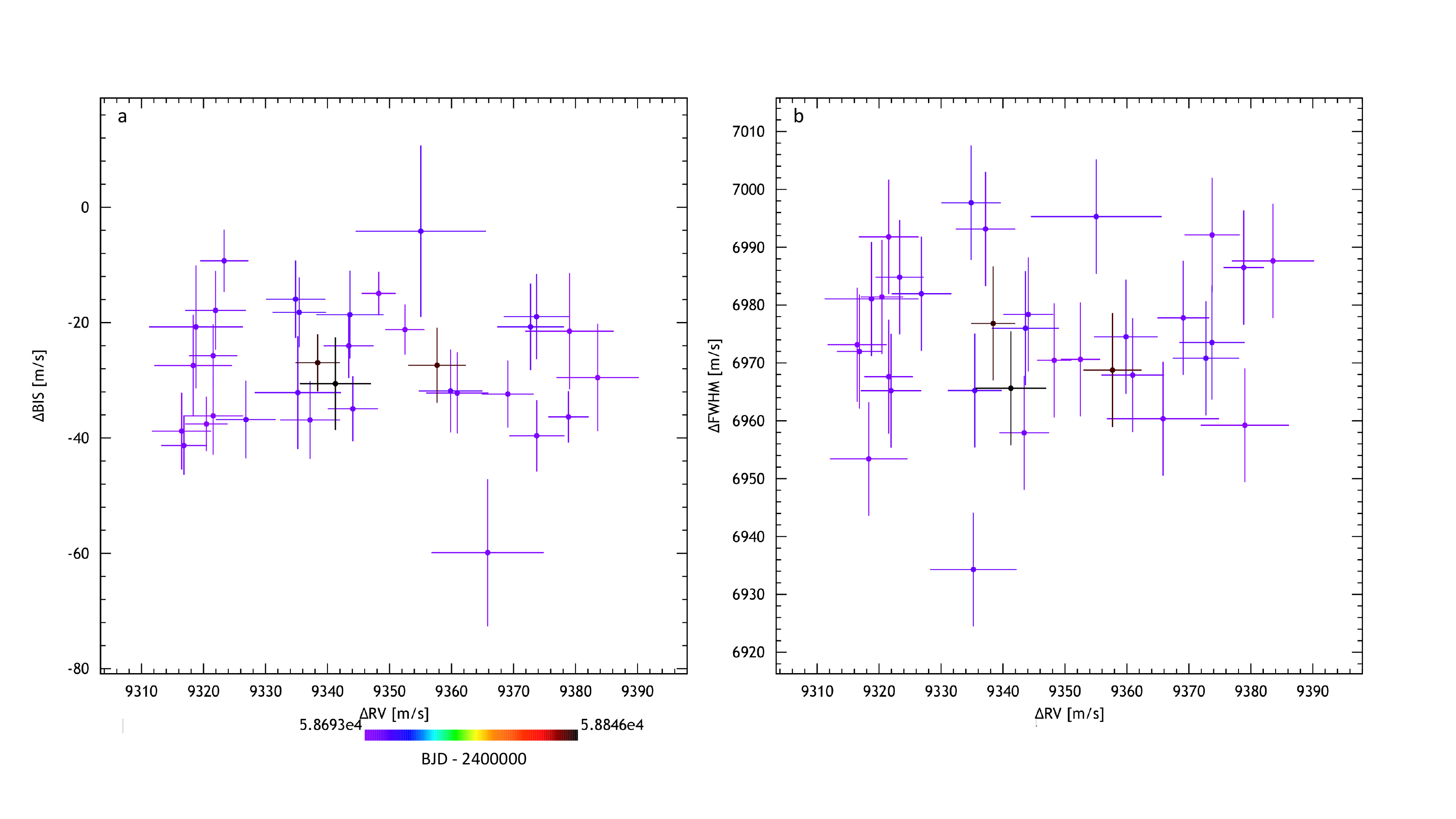}}
\caption{\textbf{a} HARPS radial velocities plotted against their bisector value. Colours represent time of observation measured in BJD-2400000. \textbf{b} as a for the full-width-half-maximum of the CCF. No correlation is seen in either case.}
\label{}
\end{figure}

\begin{figure}
\resizebox{\hsize}{!}{\includegraphics{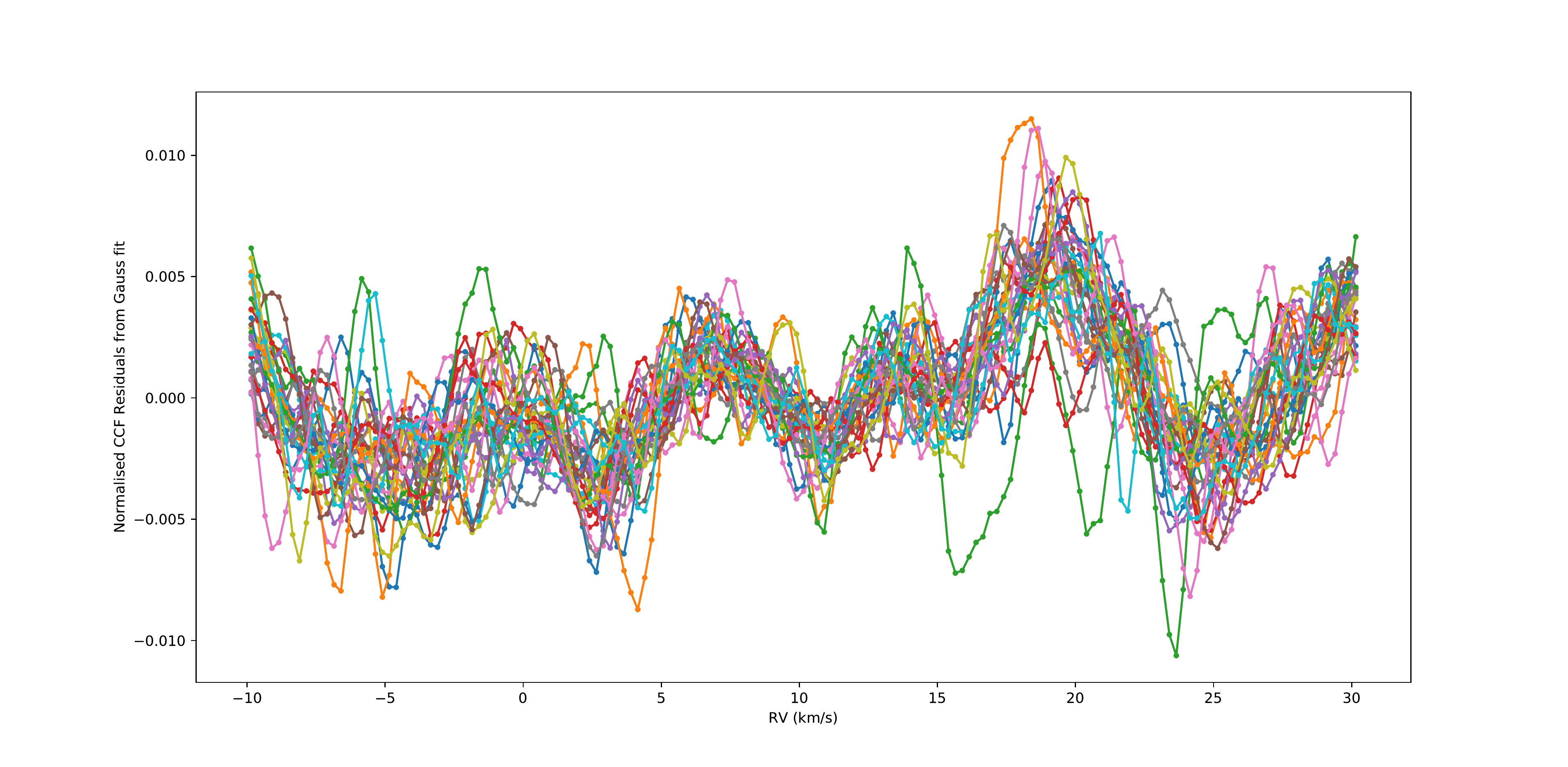}}
\caption{The CCFs of each of the HARPS spectra computed using a G2V template. A gaussian fit has been removed to leave the residual noise. No clear evidence for a contaminating star is seen.}
\label{}
\end{figure}


\begin{figure}
\resizebox{\hsize}{!}{\includegraphics{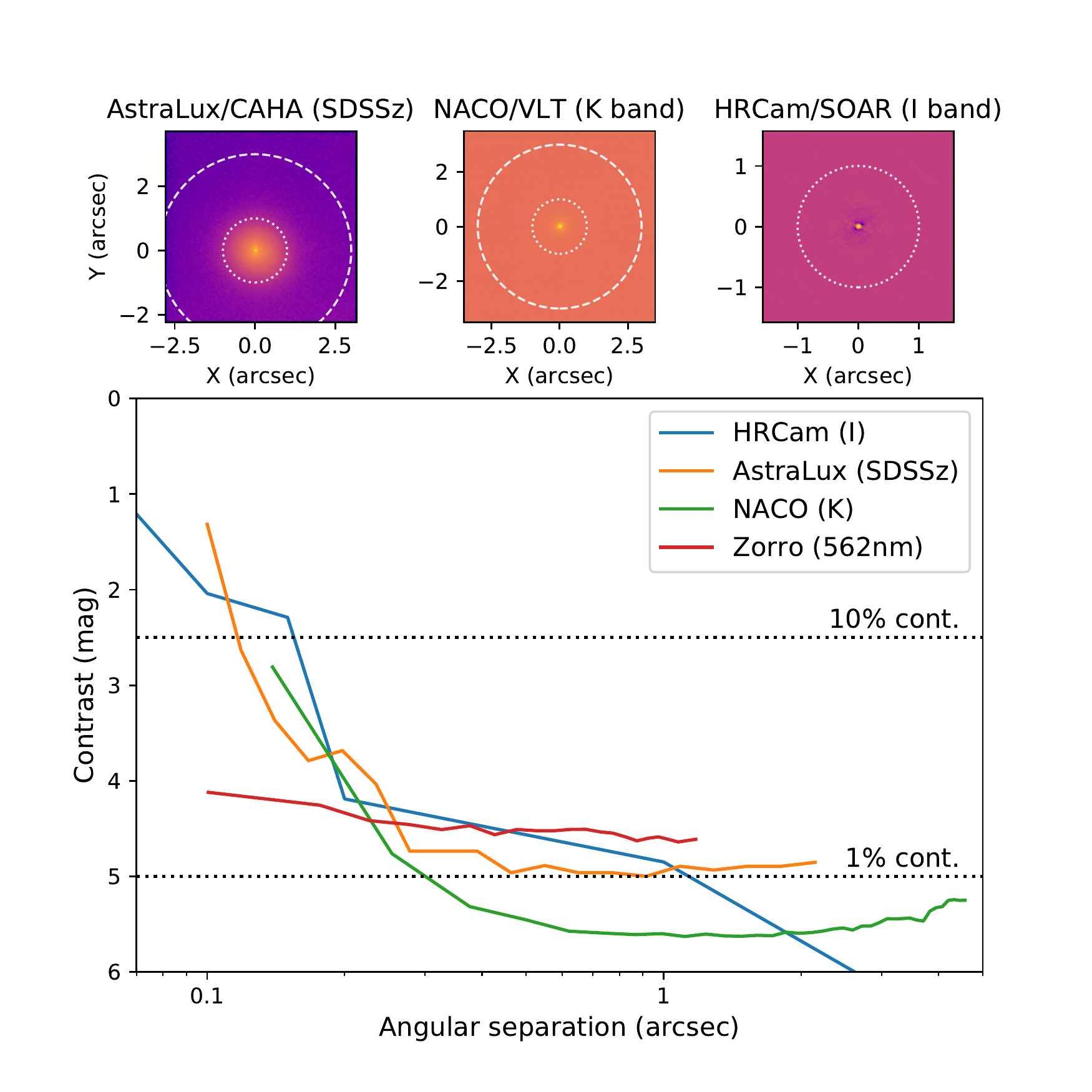}}
\caption{Collected high-resolution imaging from AstraLux/CAHA, NaCo/VLT, HRCam/SOAR and Zorro. The images are shown at top for AstraLux, NaCo and HRCam and sensitivity curves for all four below. 1\% and 10\% contrast curves are plotted.}
\label{}
\end{figure}

\begin{figure}
\resizebox{\hsize}{!}{\includegraphics{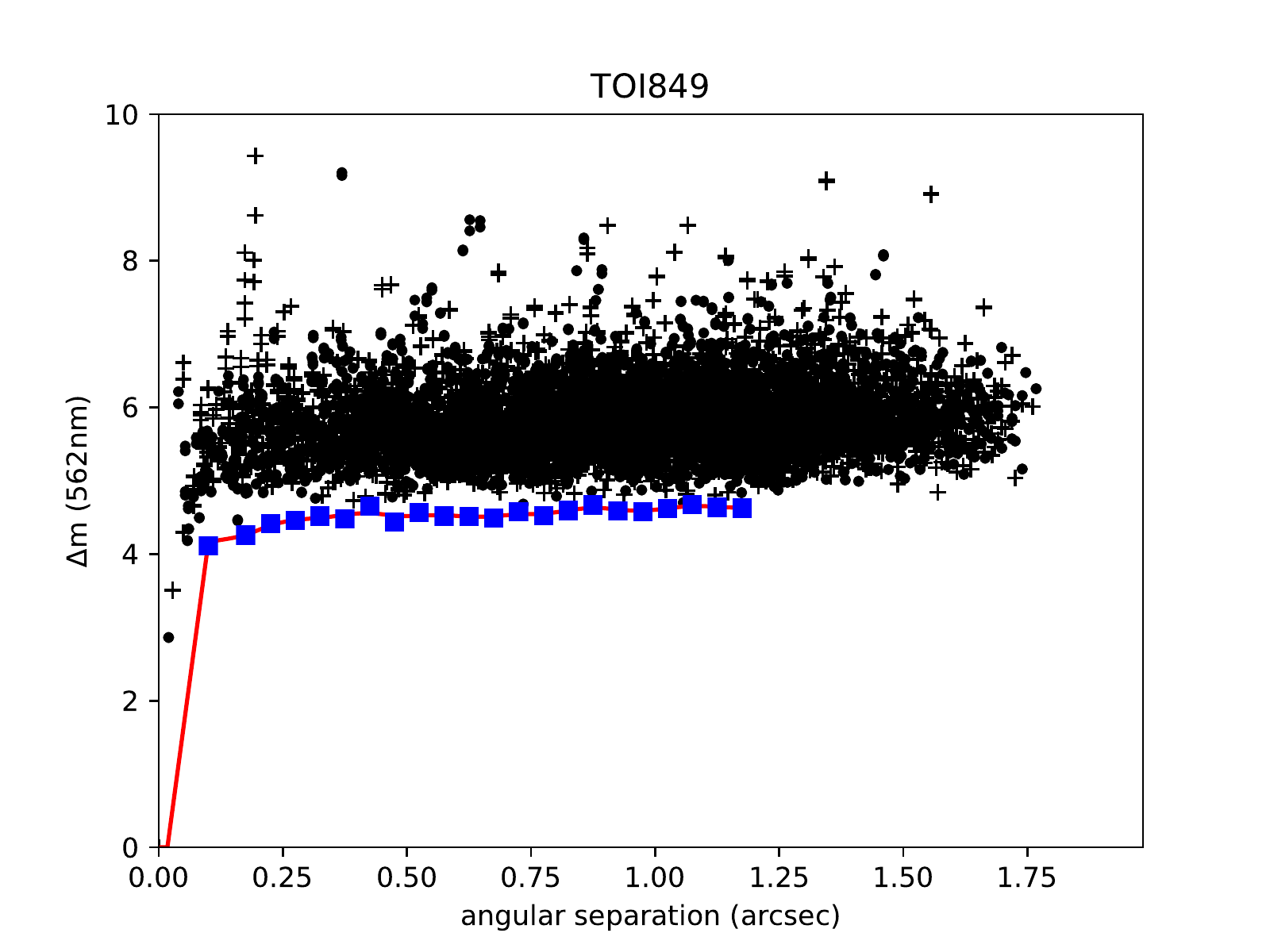}}
 \caption{Zorro speckle observation of TOI-849 taken at 562\,nm. Our simultaneous 832\,nm observation provides a similar result. The red line fit and blue points represent the $5\sigma$ fit to the sky level (black points) revealing that no companion star is detected from the diffraction limit (17\,mas) out to 1.75\,\arcsec within a $\Delta$\,mag of 5 to 6.}
\label{}
\end{figure}

\begin{figure}
\resizebox{\hsize}{!}{\includegraphics{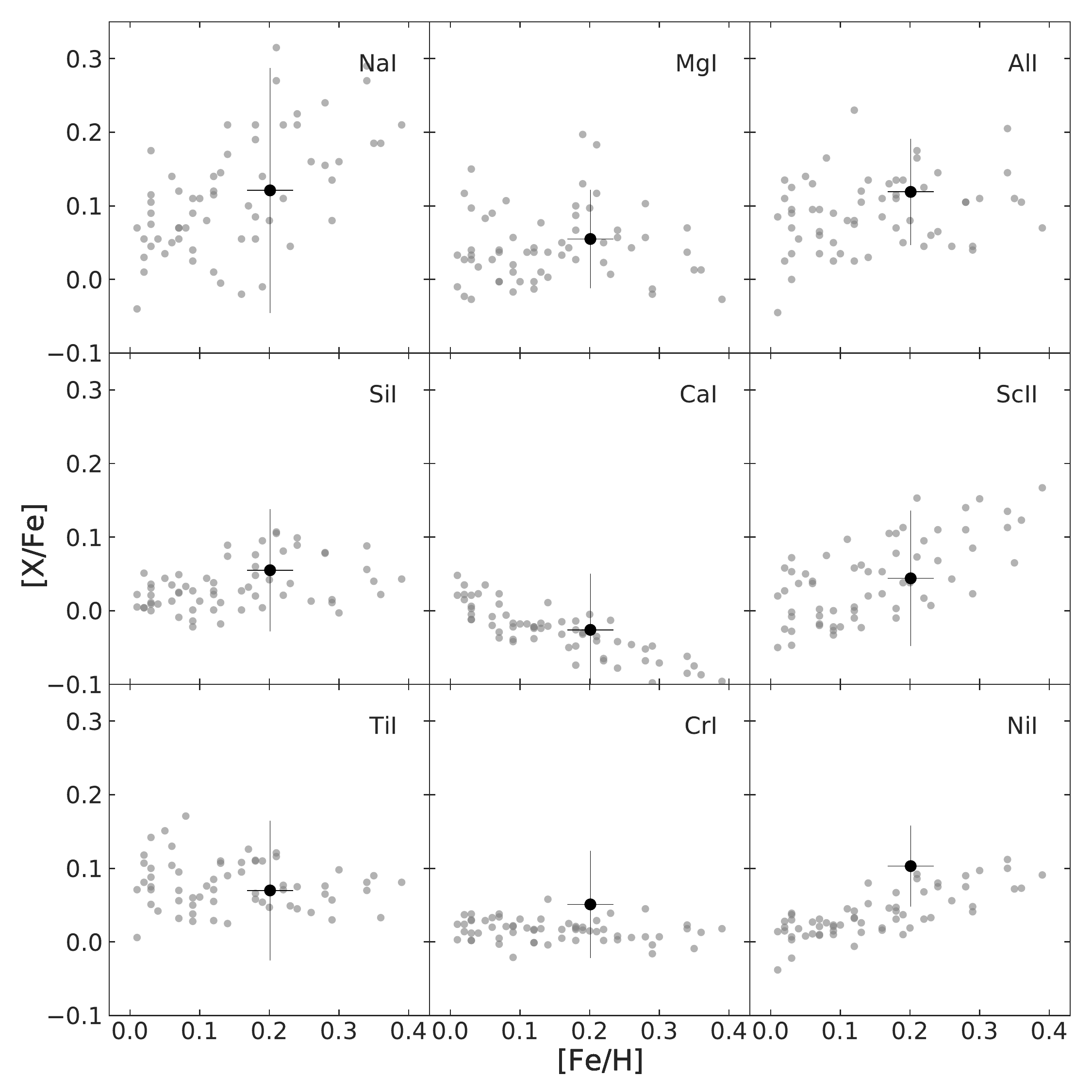}}
\caption{Abundance ratio [X/Fe] against stellar metallicity for TOI-849 (black) and for the field stars from the HARPS sample (gray) with similar stellar parameters: T$_{eff}$= 5329$\pm$200 K, $\log{g}$= 4.28$\pm$0.20 dex,  and [Fe/H]= +0.20$\pm$0.20 dex.}
\label{}
\end{figure}

\begin{figure}
\resizebox{\hsize}{!}{\includegraphics{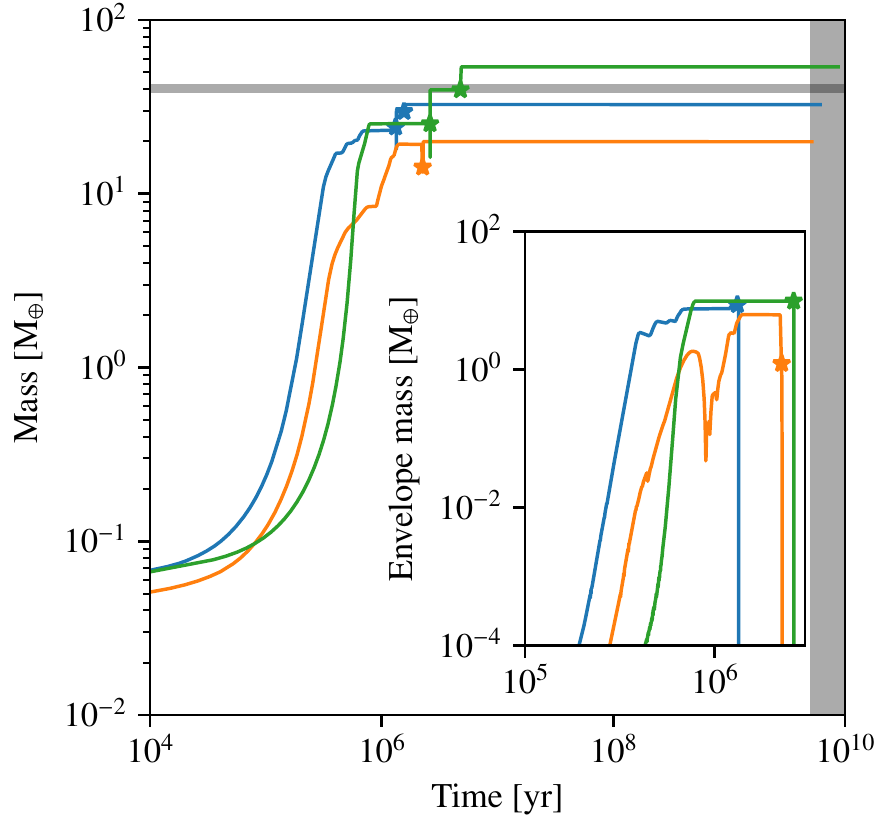}}
\caption{Planet mass against time for three similar planets to TOI-849b in the Bern Population Synthesis models. Grey shaded regions mark the parameters of TOI-849b. Stars mark the time of a giant impact. The inset shows the envelope mass, which is removed after collision.}
\label{}
\end{figure}

\renewcommand{\tablename}{Extended Data Table}
\setcounter{table}{0}

\begin{table}
\caption{}
\label{}
\begin{tabular}{llllllllr}
\hline
\hline
BJD & RV & $\sigma_\textrm{RV}$ & CCF FWHM & CCF Contrast & Bisector & S/N(50) & T$_\textrm{exp}$ & Airmass$_\textrm{start}$ \\
$d$ & $kms^{-1}$ &  $kms^{-1}$ & $kms^{-1}$ &  & $kms^{-1}$ & & s &  \\
\hline
2458692.78910182 & 9.320464 & 0.003341 & 6.9814 & 57.568 & -0.0376 & 28.9 & 1800 & 1.464\\
2458692.87197814 & 9.348264 & 0.002660 & 6.9705 & 57.606 & -0.0149 & 35.0 & 1800 & 1.073\\
2458693.78193905 & 9.379034 & 0.007109 & 6.9592 & 58.265 & -0.0215 & 16.7 & 1500 & 1.49\\
2458693.86713031 & 9.383575 & 0.006590 & 6.9876 & 57.935 & -0.0295 & 17.6 & 1200 & 1.069\\
2458694.79485824 & 9.352493 & 0.003080 & 6.9706 & 57.677 & -0.0212 & 31.2 & 1500 & 1.365\\
2458694.89266609 & 9.321555 & 0.004772 & 6.9918 & 57.768 & -0.0362 & 22.3 & 1200 & 1.022\\
2458695.76161626 & 9.316472 & 0.004716 & 6.9732 & 57.804 & -0.0388 & 22.9 & 1500 & 1.633\\
2458695.8578846 & 9.318335 & 0.006212 & 6.9534 & 58.095 & -0.0274 & 18.1 & 1200 & 1.078\\
2458697.77111987 & 9.373754 & 0.004389 & 6.9921 & 57.730 & -0.0396 & 24.0 & 1800 & 1.508\\
2458697.86515415 & 9.365826 & 0.009028 & 6.9604 & 58.639 & -0.0599 & 13.5 & 1200 & 1.051\\
2458698.79553074 & 9.316836 & 0.003574 & 6.9720 & 57.644 & -0.0413 & 28.1 & 1500 & 1.296\\
2458698.86273442 & 9.321557 & 0.003851 & 6.9676 & 57.785 & -0.0257 & 26.1 & 1500 & 1.055\\
2458699.77215996 & 9.343418 & 0.003958 & 6.9579 & 57.855 & -0.0240 & 26.0 & 1500 & 1.434\\
2458699.86782619 & 9.360927 & 0.004984 & 6.9679 & 57.876 & -0.0322 & 21.5 & 1200 & 1.038\\
2458700.7860712 & 9.378852 & 0.003165 & 6.9865 & 57.620 & -0.0363 & 30.7 & 1500 & 1.321\\
2458700.86459501 & 9.369082 & 0.004117 & 6.9778 & 57.755 & -0.0324 & 24.9 & 1200 & 1.039\\
2458701.74930712 & 9.337185 & 0.004769 & 6.9931 & 57.629 & -0.0369 & 22.8 & 1200 & 1.573\\
2458701.82063133 & 9.321939 & 0.004835 & 6.9652 & 57.948 & -0.0179 & 21.9 & 1200 & 1.139\\
2458701.91235041 & 9.318780 & 0.007543 & 6.9811 & 58.274 & -0.0207 & 15.3 & 1200 & 1.0\\
2458702.75424659 & 9.323328 & 0.003815 & 6.9848 & 57.651 & -0.0093 & 26.7 & 1200 & 1.504\\
2458702.82285066 & 9.344083 & 0.003979 & 6.9784 & 57.701 & -0.0349 & 25.8 & 1200 & 1.125\\
2458705.75330754 & 9.335210 & 0.006936 & 6.9343 & 57.874 & -0.0321 & 16.5 & 1200 & 1.443\\
2458705.8276873 & 9.326840 & 0.004757 & 6.9820 & 57.743 & -0.0368 & 22.3 & 1200 & 1.087\\
2458705.92257763 & 9.359851 & 0.005083 & 6.9745 & 57.812 & -0.0318 & 21.9 & 1200 & 1.007\\
2458706.89905173 & 9.373723 & 0.005230 & 6.9735 & 57.873 & -0.0190 & 20.6 & 1800 & 1.0\\
2458707.74440581 & 9.372752 & 0.005320 & 6.9708 & 57.367 & -0.0207 & 20.7 & 1800 & 1.508\\
2458707.85009529 & 9.355053 & 0.010510 & 6.9953 & 57.194 & -0.0042 & 12.4 & 1800 & 1.036\\
2458708.72594834 & 9.334852 & 0.004742 & 6.9977 & 57.107 & -0.0159 & 22.9 & 1200 & 1.626\\
2458708.82117413 & 9.335439 & 0.004273 & 6.9652 & 57.287 & -0.0182 & 24.4 & 1200 & 1.084\\
2458708.92270817 & 9.343619 & 0.005383 & 6.9760 & 57.703 & -0.0186 & 20.9 & 1200 & 1.013\\
\hline
\end{tabular}
\end{table}

\end{document}